\documentstyle[epsf,makeidx]{lamuphys}
\def\beqa{\begin{eqnarray}}
\def\enqa{\end{eqnarray}}
\def\beq{\begin{equation}}
\def\enq{\end{equation}}
\def\sn{{\rm sn}}
\def\cn{{\rm cn}}

\def\sign{{\rm sign}}
\def\nonum{\nonumber}
\makeatletter
\let\chapter\hid@chapter
\makeatother
\makeindex
\begin{document}
\pagenumbering{arabic}
\title{Thermodynamical Bethe Ansatz and Condensed Matter}

\author{ Minoru TAKAHASHI }

\institute{Institute for Solid State Physics, University of Tokyo, \\
Roppongi, Minato-ku, Tokyo 106, Japan}

\maketitle

\begin{abstract}
The basics of the thermodynamic Bethe ansatz equation are given. 
The simplest case is repulsive delta function bosons, the thermodynamic 
equation contains only one unknown function. 
We also treat the XXX model with 
spin 1/2 and the XXZ model and the XYZ model.  
This method is very useful for the 
investigation of the low temperature thermodynamics of solvable systems. 
\end{abstract}

\chapter{Introduction}
Thermodynamic Bethe ansatz equations were first introduced by Yang and Yang 
for the model of repulsive delta-function bosons. 
Later this method was extended to the other Bethe ansatz soluble models 
such as the XXX model, the XXZ model, the XYZ model, 
delta-function fermions, the Hubbard model, the t-J model and so on.  
In these theories the 0 of strings and  holes 
play an essential role. 
The number of kinds of strings is generally infinite except in some 
special cases. We can construct a set of non-linear integral equations for 
these models. These are very useful for the investigation of the 
low-temperature  thermodynamics of soluble models. 
Nowadays thermodynamic Bethe ansatz equations have been derived for 
almost all solvable one-dimensional quantum models.\par

In chapter 2, Yang and 
Yang's theory for bosons is introduced. 
Here the number of unknown functions is 
1. In chapter 3 we treat the XXX model. Here the number of unknown functions is 
infinite. But we can treat this set of integral equations analytically in 
some special cases such as the infinite temperature limit, the 
zero temperature limit, 
and the high magnetic field limit. In chapter 3 we treat the 
XXZ model. The case 
$|\Delta|>1$ is almost identical to the XXX model ($\Delta=1$). On the other 
hand the case $|\Delta|<1$ is more complicated, the number of kinds of 
strings is dependent on a parameter. If $(\cos^{-1}\Delta)/\pi$ is a rational 
number, the number of kinds of strings is finite and thermodynamic 
Bethe ansatz equations 
contain a finite number of unknown functions. 
In chapter 5 we treat the XYZ model in zero 
magnetic field. Here, the number of unknown functions is also 
dependent on parameters. The equations 
are obtained by a slight modification of XXZ model at $|\Delta|<1$. 
In chapter 6 we review numerical calculations of these Bethe ansatz 
equations.

\chapter{Repulsive delta-function bosons}

\section{Bethe ansatz equations and uniqueness of the solution}
Here we consider the system
\begin{equation}
{\cal H}=-\sum^N_{i=1}{\partial^2\over \partial x_i^2}+
2c\sum_{i<j}\delta(x_i-x_j),\label{eq:del1}
\end{equation}
the problem of bosons interacting via a repulsive
delta function potential with periodic boundary conditions. 
Lieb and Liniger showed that this model is solvable by Bethe 
ansatz and calculated the ground state energy and elementary 
excitations (\cite{lieb63a}, \cite{lieb63b}).
The wave function is assumed to be a linear combination of $N!$ 
plane waves. 
One must determine $N!$ coefficients $A(P)$ with eigenvalue $E$ at 
$x_1\le x_2\le ...\le x_N$
\beqa
&&f=\sum_PA(P)\exp[i(k_{P1}x_1+k_{P2}x_2+...+k_{PN}x_N)],\nonum\\
&&A(P)=C\epsilon(P)\prod_{j<k}(k_{Pj}-k_{Pk}+ic),~~\nonum\\
&&E=\sum_{j=1}^Nk_j^2.
\label{eq:delbwf}
\enqa
Here $P$'s are permutations and $\epsilon (P)$ is the parity of permutation 
$P$. The total momentum is given by
$$K=\sum^N_{j=1}k_j.$$
$k_j$'s are called the quasi momenta. If we apply 
periodic boundary conditions, $k_j$'s should be solutions of the 
following coupled equations
\beq
e^{ik_j L}=-\prod^N_{l=1}({k_j-k_l+ic\over k_j-k_l-ic}).
\enq
The logarithm of these equations is
\begin{equation}
k_jL=2\pi I_j-\sum_l 2\tan^{-1}({k_j-k_l \over c}).
\label{eq:delbt1}
\end{equation}
Here the $I_j$'s form a set of distinct but otherwise arbitrary 
integers (half-odd integers) for odd(even) $N$. The
quasi-momenta $k_j$ are determined by eq. (\ref{eq:delbt1}) if a set of
quantum numbers $I_j$ is given. 
The ground state energy is obtained from a Fredholm type integral equation
(\cite{lieb63a,lieb63b}).
For this model Yang and Yang found a non-linear integral equation
which gives the free energy at a given temperature (\cite{yang69}). 
This result is important
because it forms the basic theory for thermodynamics of other solvable
models.
\par
At first we should show that the solution of 
equation (\ref{eq:delbt1}) is unique for 
a given set of $I_j$'s. 
The equations (\ref{eq:delbt1}) are equivalent to the following equations
\beqa
&&{\partial B(k_1,k_2,...,k_N)\over \partial k_j}=0, \nonum\\
&&B(k_1,k_2,...,k_N)\equiv
{L\over 2}\sum k_j^2-2\pi I_jk_j+\sum_{j<l}\theta_1(k_j-k_l),\nonum\\
&&\theta_1(x)=x\tan^{-1}(x/c)-{c\over 2}\ln(1+(x/c)^2).
\enqa
The extremal point of the function $B$ with $N$ variables is determined by 
the solution of (\ref{eq:delbt1}).  It gives an eigenstate
of the Hamiltonian. Let us consider the following $N\times N$ matrix
\beq
B_{jl}={\partial^2 B\over \partial k_j\partial k_l}
=\delta_{jl}(L+\sum_m{2c\over c^2+(k_j-k_m)^2})-{2c\over c^2+(k_j-k_l)^2}.
\enq
This matrix is always positive definite because
\beq
\sum_{lj}u_lB_{lj}u_j=L\sum u_l^2
+\sum_{l<j}{2c\over c^2+(k_j-k_l)^2}(u_j-u_l)^2\ge 0,
\enq
for an arbitrary real vector $\{u_j\}$. The function $B$ is a concave 
function in
$N$ dimensional space. Thus a solution of equation (\ref{eq:delbt1}) for a given
set of $\{I_j\}$ is unique.

\section{Holes of quasi momenta and their distribution function}
In the limit of $c\to \infty$ this set of eigenstates
is complete. Let us consider the following function
\beq
h(k)=k+{2\over L}\sum_j\tan^{-1}{k-k_j\over c}.
\enq
The position of a hole is defined by
$$h(k_h)={2\pi\over L}\times{\rm unoccupied ~~(half-odd)~~ integer}.$$
We define the distribution function of holes $\rho^h(k)$ such that the 
number of holes between $k$ and $k+dk$ is $\rho^h(k)Ldk$ and the density 
of particles $\rho(k)$ such that the number of particles between $k$ 
and $k+dk$ is $\rho(k)Ldk$.
\index{hole distribution function}
\index{particle distribution function}
In the thermodynamic limit we have
\beq
2\pi\int^k\rho(t)+\rho^h(t)dt=h(k)=k+2\int \tan^{-1}{k-k'\over c}\rho(k')dk'.
\enq
Differentiating this yields
\beq
2\pi(\rho(k)+\rho^h(k))=1+2\int^\infty_{-\infty}{c\rho(k')dk'
\over c^2+(k-k')^2}.
\label{eq:delbteq2}
\enq
From equation (\ref{eq:delbwf}) one obtains the energy and the particle number 
per unit length
\beq
e=\int^\infty_{-\infty}k^2\rho(k)dk,\quad n=\int^\infty_{-\infty}\rho(k)dk.
\enq
The entropy of the distribution between $k$ and $k+dk$ is the logarithm of
number of orderings of $L\rho(k)dk$ particles and $L\rho^h(k)dk$ holes
\beqa
\ln{[L(\rho(k)+\rho^h(k))dk]!\over [L\rho(k)]![L\rho^h(k)]!}
&&=Ldk[(\rho(k)+\rho^h(k))\ln(\rho(k)+\rho^h(k))\nonum\\
&&-\rho(k)\ln\rho(k)-
\rho^h(k)\ln\rho^h(k)].\nonum
\enqa
\index{Yang-Yang theory}\index{Stirling formula}
Here we use the Stirling formula $\ln( n!)\simeq n(\ln n-1)$. Then the 
entropy per unit length is
\beq
{\it s}=\int^\infty_{-\infty}
(\rho(k)+\rho^h(k))\ln(\rho(k)+\rho^h(k))-\rho(k)\ln\rho(k)-
\rho^h(k)\ln\rho^h(k)dk.
\label{eq:delbent}
\enq

\section{Thermodynamic equilibrium}
The free energy per unit length $f=e-T{\it s}$ must be 
minimized under the condition
that $n$ is constant. $f$ and $n$ are functionals of $\rho(k)$
and $\rho^h(k)$. Thus we should determine $\rho(k)$
and $\rho^h(k)$ to minimize $f-An$. Next we look
for parameter $A$ such that $n$ takes on its required value. In variational
calculus $A$ is called a Lagrange multiplier. \index{variational calculation}
\index{Lagrange multiplier}
At the minimum point,
the variation of $f-An$ must be zero for any infinitesimal variation of
functions
\beqa
&&0=\delta \int^\infty_{-\infty}
(k^2-A)\rho(k)-T\bigl\{(\rho(k)+\rho^h(k))\ln(\rho(k)+\rho^h(k))\nonum\\
&&-\rho(k)\ln\rho(k)-\rho^h(k)\ln\rho^h(k)\bigr\} dk\nonum\\
&&=\int\Bigl\{k^2-A-T\ln\bigl({\rho(k)+\rho^h(k)\over\rho(k)}\bigr)\bigr\}
\delta\rho(k)-T\ln\bigl({\rho(k)+\rho^h(k)\over \rho^h(k)}\bigr)
\delta\rho^h(k) dk.\nonum\\
\label{eq:xt8}
\enqa
One should note that $\delta\rho(k)$ and $\delta\rho^h(k)$ are not independent.
From (\ref{eq:delbteq2})
we have
$$\delta\rho^h(k)=-\delta\rho(k)+{1\over \pi}\int^\infty_{-\infty}
 {c\delta\rho(k')dk'\over c^2+(k-k')^2}.$$
Substituting this into (\ref{eq:xt8}), one obtains
\beqa
&&0=\int^\infty_{-\infty}\delta\rho(k)\nonum\\
&&\times\bigl\{
k^2-A-T\ln({\rho^h(k)\over\rho(k)})
-{T\over \pi}\int^\infty_{-\infty}
dq{c\over c^2+(k-q)^2}\ln(1+{\rho(q)\over \rho^h(q)})dq\bigr\}.
\enqa
This equation must hold for any arbitrary infinitesimal change of the 
function $\rho(k)$.
Thus
\beq
T\ln({\rho^h(k)\over \rho(k)})=k^2-A-{T\over \pi}\int^\infty_{-\infty}
dq{c\over c^2+(k-q)^2}\ln(1+{\rho(q)\over \rho^h(q)})dq,
\enq
must be satisfied at thermodynamic equilibrium.
If we put 
\par\noindent
$\epsilon(k)=T\ln(\rho^h(k)/\rho(k))$, this becomes
\beq
\epsilon(k)=k^2-A-{T\over \pi}\int^\infty_{-\infty}
dq{c\over c^2+(k-q)^2}\ln(1+\exp(-{\epsilon(q)\over T}))dq.
\label{eq:delbepsi}
\enq
$\epsilon(k)$ can be determined by iteration. 
This function has the physical meaning as the excitation energy 
for an elementary excitation, as will be shown in next section. 
From equation (\ref{eq:delbteq2}) we have
\beq
2\pi\rho(k)(1+\exp({\epsilon(k)\over T}))=1+2c\int^\infty_{-\infty}
{\rho(q)dq\over c^2+(k-q)^2}.
\label{eq:xt12}
\enq
From this equation we can determine $\rho(k)$. \par
If equation (\ref{eq:delbepsi}) is differentiated with respect to 
the chemical potential A, one obtains a linear integral equation
\beq
{\partial \epsilon(k,A)\over \partial A}
=-1+\int dq{c\over c^2+(k-q)^2}{1\over 1+\exp({\epsilon(k)\over T})}
{\partial \epsilon(q,A)\over \partial A}.
\enq
Comparing this with (\ref{eq:xt12}) one finds that 
\beq
{\partial \epsilon(k,A)\over \partial A}
=-2\pi\rho(k)(1+\exp({\epsilon(k)\over T}))=-2\pi(\rho(k)+\rho^h(k)).
\enq
The entropy density (\ref{eq:delbent}) can be written
$${\it s}=\int[(\rho+\rho^h)\ln(1+\exp(-\epsilon(k)/T))
+\rho(k)\epsilon(k)/T]dk.$$
The free energy density is
\beq
f=\int(k^2-\epsilon(k))\rho(k)-T(\rho(k)+\rho^h(k))
\ln(1+\exp(-{\epsilon(k)\over T}))dk.
\label{eq:feb}
\enq
Substituting (\ref{eq:delbepsi}) into (\ref{eq:feb}) one obtains
\beqa
&&f=\int[A+{T\over \pi}\int{dq c\over c^2+(k-q)^2}
\ln(1+\exp(-{\epsilon(q)\over T}))]
\rho(k)\nonum\\
&&-T(\rho(k)+\rho^h(k))\ln(1+\exp(-{\epsilon(k)\over T}))dk\nonum\\
&&=\int A\rho(k)dk-T\int\ln(1+\exp(-{\epsilon(k)\over T}))\nonum\\
&&\times[\rho(k)+\rho^h(k)-{1\over \pi}\int
{c\over c^2+(k-q)^2}\rho(q)dq]dk.
\enqa
Substituting (\ref{eq:delbteq2}) we have a very simple expression for the 
free energy density
\beq
f=An-T\int{dk\over 2\pi}\ln(1+\exp(-{\epsilon(k)\over T})).
\enq
The thermodynamic potential density $g=f-An$ is
\beq
g(T,A)=-T\int\ln(1+\exp(-{\epsilon(k)\over T})){dk\over 2\pi}.
\label{eq:delbpot}
\enq
\index{thermodynamic potential}
\par\noindent
For a given temperature $T$ and chemical potential $A$, 
$\epsilon(k)$ is determined through the non-linear integral
equation (\ref{eq:delbepsi}) and obtain
thermodynamic potential $g$ from (\ref{eq:delbpot}).
All thermodynamic quantities are derived from $g(T,A)$ through thermodynamic
relations
\beq
n=-{\partial g\over\partial A},\quad {\it s}=-{\partial g\over \partial T},
\quad e=g+An+T{\it s},....
\enq
\index{thermodynamic relations}
The pressure $p$ is $-g$.

\section{Elementary excitations}
We consider the change of energy and momentum when one particle 
is moved from the thermodynamic equilibrium. 
Assume that the $l$-th particle is removed and one particle is added between the
$m$-th and $m+1$-th particle
\beqa
&&\{I_1>I_2>I_3>...>I_N\}\to\nonum\\
&&\{I_1>I_2>...>I_m>I'_l>I_{m+1}...I_{l-1}>I_{l+1}>...>
I_{N}\}\nonum
\enqa
The Bethe ansatz equations for the original state are
\beq
Lk_j=2\pi I_j-\sum_{n=1}^N2\tan^{-1}{k_j-k_n\over c}~~j=1,...,N.
\label{eq:bosex1}
\enq
For the excited state they are
\beqa
&&Lk'_j=2\pi I_j-2\tan^{-1}{k'_j-k'_l\over c}
-\sum_{n\ne m}2\tan^{-1}{k'_j-k'_n\over c},~~j\ne l\nonum\\
\label{eq:bosex2}\\
&&Lk'_l=2\pi I'_l-\sum_{n\ne m}2\tan^{-1}{k'_l-k'_n\over c}.\label{eq:bosex3}
\enqa
The change of momentum and energy are
\beqa
\Delta K=k'_l-k_l+\sum_{n\ne l}(k'_n-k_n),\nonum\\
\Delta E=k'^2_l-k_l^2+\sum_{n\ne l}(k'^2_n-k_n^2)
\enqa 
Subtracting (\ref{eq:bosex1}) from (\ref{eq:bosex2}) we have 
\beqa
&&\Delta k_jL+\sum{2c\over c^2+(k_j-k_l)^2}(\Delta k_j-\Delta k_l)\nonum\\
&&=2\tan^{-1}{k_j-k_l\over c}-2\tan^{-1}{k_j-k'_l\over c}.
\enqa
Then equation for the back-flow $J(k)=\rho(k)\Delta k L$ is
\beqa
&&(1+{\rho^h(k)\over\rho(k)})J(k;k_l',k_l)-\int {c\over \pi(c^2+(k-k')^2)}J(k';k_l',k_l)dk'\nonum\\
&&={1\over \pi}(\tan^{-1}{k-k_l\over c}-\tan{k-k'_l\over c}). 
\enqa
Energy and momentum changes are
\beqa
&&\Delta K(k_l',k_l)=k_l'-k_l+\int J(k;k_l',k_l) dk,\nonum\\
&&\Delta E(k_l',k_l)=k_l^{'2}-k_l^2+\int 2kJ(k;k_l',k_l) dk.
\enqa
It is clear that $\Delta K(k_l,k_l)=\Delta E(k_l,k_l)=0$. The differentiation 
of these with respect to $k_l'$ is
\beqa
&&{\partial \Delta K\over \partial k_l'}=1+\int u(k;k_l')dk\nonum\\
&&{\partial \Delta E\over \partial k_l'}=2k_l'+\int 2k u(k;k_l')dk,
\label{eq:elmqe}
\enqa
where 
\beqa
&&u(k;k_l')\equiv  {\partial \over \partial k_l'}J(k;k_l',k_l),\nonum\\
&&(1+\eta(k))u(k;k_l')-\int a(k-k')u(k';k_l')dk'=a(k-k_l').\label{eq:ueqa}
\enqa
Using this integral equation we find that $u(k;k_l')$ is given by the 
infinite series
\beqa
&&u(k;k_l')={1\over 1+\eta(k)}
\Bigl[a(k-k_l')+\int dk_1 {a(k-k_1)\over 1+\eta(k_1)}
a(k_1-k_l')\nonum\\
&&+\int\int dk_1dk_2{a(k-k_1)\over 1+\eta(k_1)}
{a(k_1-k_2)\over 1+\eta(k_2)}a(k_2-k_l')+...\Bigr].
\enqa
Substituting this into (\ref{eq:elmqe})  we get an expression of  
${\partial \Delta K\over \partial k_l'}$ and 
${\partial \Delta E\over \partial k_l'}$ as an infinite series
\beqa
&&{\partial \Delta K\over \partial k_l'}=1+
\int dk_1 a(k_l'-k_1){1\over 1+\eta(k_1)}\nonum\\
&&+\int\int dk_1dk_2 a(k_l'-k_1){1\over 1+\eta(k_1)}
a(k_1-k_2){1\over 1+\eta(k_2)}+...\nonum\\
&&{\partial \Delta E\over \partial k_l'}=2k_l'+
\int dk_1 a(k_l'-k_1){1\over 1+\eta(k_1)}2k_1\nonum\\
&&+\int\int dk_1dk_2 a(k_l'-k_1){1\over 1+\eta(k_1)}
a(k_1-k_2){1\over 1+\eta(k_2)}2k_2+...\nonum
\enqa
From these infinite series linear integral equations for 
${\partial \Delta K\over \partial k_l'}$ and 
${\partial \Delta E\over \partial k_l'}$ are obtained,
\beqa
{\partial \Delta K(k_l',k_l)\over \partial k_l'}=
1+\int a(k_l'-k_1){1\over 1+\eta(k_1)}
{\partial \Delta K(k_1,k_l)\over \partial k_1},\nonum\\ 
{\partial \Delta E(k_l',k_l)\over \partial k_l'}=
2k_l'+\int a(k_l'-k_1){1\over 1+\eta(k_1)}
{\partial \Delta E(k_1,k_l)\over \partial k_1}.
\enqa
On the other hand we have equations for $2\pi(\rho(k)+\rho^h(k))$ 
and $\partial\epsilon(k)/\partial k$ from (\ref{eq:delbteq2}) and 
(\ref{eq:delbepsi})
\beqa
&&2\pi(\rho(k)+\rho^h(k))=1+\int a(k-k')
{2\pi(\rho(k')+\rho^h(k'))\over 1+\eta(k')}dk',\nonum\\
&&{\partial\epsilon(k)\over \partial k}=2k+\int a(k-k'){1\over 1+\eta(k')}
{\partial\epsilon(k')\over \partial k'}dk'.
\enqa
The solution of non-singular linear integral equation is unique. 
Thus one obtains
\beqa
&&{\partial \Delta K(k_l',k_l)\over \partial k_l'}=
2\pi(\rho(k_l')+\rho^h(k_l')),\nonum\\
&&{\partial \Delta E(k_l',k_l)\over \partial k_l'}=
{\partial\epsilon(k'_l)\over \partial k'_l}.
\enqa
Integrating these we have
\beqa
&&\Delta K(k_l',k_l)=2\pi \int^{k_l'}_{k_l}\rho(k)+\rho^h(k)dk,\nonum\\
&&\Delta E(k_l',k_l)=\epsilon(k_l')-\epsilon(k_l).
\enqa
Thus $\epsilon(k)=T\ln(\rho^h(k)/\rho(k))$ has 
the physical meaning as the energy of the elementary excitation.

\section{Some special limits}
\subsection{$c=\infty$ limit}
In the thermodynamic limit the system is equivalent to ideal 
spinless fermions 
\beqa
&&p={1\over 2\pi}\int {k^2dk\over \exp(k^2-A)/T+1}=\nonum\\
&&-{T\over 2\pi}\int \ln(1+\exp[(A-k^2)/T])dk=\lim_{L\to \infty} -G/L,
\label{eq:delpres}
\enqa
by the partial differentiation with respect to $k$.
In this limit $c/(c^2+(k-q)^2)$ in the integrand of
(\ref{eq:delbepsi}) is zero. Then we
have have very simple solution $\epsilon(k)=k^2-A$. 
From equation (\ref{eq:xt12}) we have
\beq
\rho(k)={1\over 2\pi}{1\over \exp[(k^2-A)/T]+1}.
\enq
From (\ref{eq:delbpot}) the thermodynamic potential per unit length is
\beq
g=-{1\over 2\pi}\int dk\ln(1+\exp(-{k^2-A\over T})).
\enq
This is equivalent to (\ref{eq:delpres}). \par\noindent
\subsection{$c={\it 0}+$ limit}
In this limit integration kernel $c/(c^2+(k-q)^2)$ can be replaced by
$\pi\delta(k-q)$. Then (\ref{eq:delbepsi}) becomes
$\epsilon(k)=k^2-A-T\ln(1+\exp(-\epsilon(k)/T))$. Then we obtain
\beq
\epsilon(k)=T\ln(\exp((k^2-A)/T)-1).
\enq
The Gibbs free energy per site $g$, pressure $p$, $\rho(k)$ and $\rho^h(k)$ are
\beq
g=-p=T\int\ln(1-\exp(-{k^2-A\over T})){dk\over 2\pi},\quad
\rho^h(k)={1\over 2\pi},$$
$$\rho(k)={1\over 2\pi}{1\over \exp((k^2-A)/T)-1}.
\enq
This result coincide with that for ideal bosons. 
\par\noindent

\subsection{$T={\it 0}+$ limit}
Generally speaking $\epsilon(k)$ is a monotonically increasing function
of $k^2$. At $T=0+$ we assume that $\epsilon(\pm q_0)=0$. Thus we have
$$\rho(k)=0\quad{\rm for}\quad k^2>q_0^2, \quad
\rho^h(k)=0\quad{\rm for}\quad k^2<q_0^2.$$
Equations  (\ref{eq:delbteq2}) and (\ref{eq:delbepsi}) for $k<q_0$ are
\beqa
&& 2\pi\rho(k)=1+2c\int^{q_0}_{-q_0}{\rho(q)dq\over c^2+(k-q)^2},\nonum\\
&&\epsilon(k)=k^2-A+{c\over \pi}\int^{q_0}_{-q_0}
{\epsilon(q)dq\over c^2+(k-q)^2}.
\enqa
The first equation is equivalent with Lieb-Liniger equation(\cite{lieb63a}),
\begin{equation}
\rho(k)={1\over 2\pi}+\int^B_{-B}{c/\pi\over c^2+(k-q)^2}\rho(q)dq.
\label{eq:delbl&l}
\end{equation}
\index{Lieb-Liniger equation}
\par\noindent
The above theory was introduced by Yang-Yang (\cite{yang69}).
C.P. Yang solved numerically this equation(\cite{pyan70}). Very surprisingly 
it seems that this simple non-linear equation gives the exact free energy 
in the thermodynamic limit of the 1D repulsive Bosons. \index{thermodynamic 
Bethe ansatz equation}
The next problem is to
find thermodynamic Bethe ansatz equation for other soluble models.
As the next simplest case we treat the $S=1/2$ XXX chain.

\chapter{Thermodynamics of the XXX chain}
\section{Wave functions of the XXX chain}
The Heisenberg model was the first model to be treated by the
method of Bethe ansatz (\cite{bethe31}). \index{Bethe}\index{Heisenberg model}
\index{XXZ model}
In the beginning of the 1930's 
only the ferromagnetic case was
considered
\beqa
&&{\cal H}=-J\sum_{l=1}^NS_l^xS_{l+1}^x+S_l^yS_{l+1}^y+S_l^zS_{l+1}^z
-2h\sum_{l=1}S^z_l,\nonum\\
&& h\ge 0,~~ {\bf S}_{N+1}\equiv {\bf S}_1.
\label{eq:xxxham}
\enqa
This Hamiltonian is defined on a $2^N$ dimensional vector space. The
space is classified by the total $S^z=\sum S^z_l$.
The ground state is the state where all spins are up and $S^z=N/2$
\begin{equation}
{\cal H}\vert 0>=E_0\vert 0>,\quad E_0=-JN/4-Nh.
\end{equation}   
Write a general state $|\Psi >$ in terms of a wave function $f$, 
\begin{equation}
|\Psi>=\sum f(n_1,n_2,...,n_M)S_{n_1}^-S_{n_2}^-...S_{n_M}^-\vert 0>,
\label{eq:xxxfunc}
\end{equation}
where $1\le n_1<n_2<...<n_M\le N$ and $2M\le N$. 
The eigenvalue equation
is
\beqa
&&-{J\over 2}\sum_j(1-\delta_{n_j+1, n_{j+1}})
\Bigl\{f(n_1,...,n_j+1,n_{j+1},...,n_M)\nonum\\
&&+f(n_1,...,n_j,n_{j+1}-1,...,n_m)\Bigr\}\nonum\\
&&+\Bigl\{E_0-E+(J+2h)M-J\sum_j\delta_{n_j+1,n_{j+1}} \Bigr\}
f(n_1,n_2,...,n_M)=0.\nonum\\
\label{eq:xxxfunc2}
\enqa
Next we assume that the wave function is of the following form
\begin{equation}
f(n_1,n_2,...,n_M)=\sum_P^{M!} A(P)\exp(i\sum_{j=1}^Mk_{Pj}n_j).
\label{eq:xxxf2}
\end{equation}
Choosing 
\beqa
&&A(P)=\epsilon(P)\prod_{j<l}(e^{i(k_{Pj}+k_{Pl}}+1-2e^{ik_{Pl}}),\\
&&E-E_0=2hM+J\sum_{j=1}^M(1-\cos k_j).
\enqa
insures that $f$ satisfies (\ref{eq:xxxfunc2}). 
If we put $e^{ik_j}=(x_j+i)/(x_j-i)$, the wave function and energy are written 
as follows
\beqa
&&f(n_1,n_2,...,n_M)=\sum_P^{M!} A(P)\prod_{j=1}^M
({x_{Pj}+i\over x_{Pj}-i})^{n_j},\\
&&A(P)=D\epsilon(P)\prod_{j<l}(x_{Pj}-x_{Pl}-2i),\label{eq:xxxA}\\
&&E-E_0=2hM+2J\sum_{j=1}^M{1\over x_j^2+1}.
\enqa
The $x_j$'s are called rapidities. \index{rapidity}
The periodic boundary condition 
$$f(x_1,x_2,...,x_M)=f(x_2,x_3,...,x_M,x_1+N),$$
implies the $x_j$ satisfy
\begin{equation}
\Bigl({x_j+i\over x_j-i}\Bigr)^N=\prod_{l\ne j}
\Bigl({x_j-x_l+2i\over x_j-x_l-2i }\Bigr),~~j=1,...,M.
\label{eq:xxxal2}
\end{equation}
In terms of the rapidity variable, the periodic boundary conditions
take on a very simple form. 

\section{Hulthen's solution for the antiferromagnet}
\index{Hulthen}\label{sec:sec33}
In actual magnetic substances, the ferromagnetic case is rare. 
Usual 
one- dimensional
magnetic substances are antiferromagnetic 
(the $J<0$ case of equation (\ref{eq:xxxham})).
The logarithm of equation (\ref{eq:xxxal2})
is
\begin{equation}
2N\tan^{-1}x_j=2\pi I_j+2\sum_{l=1}^M\tan^{-1}{x_j-x_l\over 2},
\label{eq:xxxeqn}
\end{equation}
where $I_j$ is an integer(half-odd integer) for odd(even) $N-M$.
The total momentum is given by
$$K=\pi(1-(-1)^M)/2-{2\pi \over N}\sum_j I_j.$$
For simplicity we set $N$ to be even. 
One can
show that the lowest energy state in the subspace of total $S_z=N/2-M$ 
is given by
\begin{equation}
I_j=(M+1-2j)/2,\quad j=1,2,...,M.
\end{equation}
In the thermodynamic limit $x_j$'s distribute from $-B$ to $B$. From equation
(\ref{eq:xxxeqn}) we have
\begin{equation}
\tan^{-1}x=\pi\int^x\rho(t)dt+\int^B_{-B}\tan^{-1}{x-y\over 2}\rho(y)dy.
\end{equation}
Differentiating with respect to $x$ yields
\begin{equation}
\rho(x)={1\over\pi}{1\over x^2+1}-\int^B_{-B} {1\over\pi}
{2\over (x-y)^2+4}\rho(y)dy.
\label{eq:xxxfre}
\end{equation}
The energy and magnetization per site are
\begin{equation}
{E\over N}={\vert J\vert \over 4}-h
+\int^B_{-B}[2h-{2\vert J\vert \over x^2+1}]\rho(x)dx,
\label{eq:xxxenfre}
\end{equation}
\begin{equation}
{S_z\over N}={1\over 2}-\int^B_{-B}\rho(x) dx.
\label{eq:xxxmagfre}
\end{equation}
This integral equation can be solved in the case of infinite $B$.
We define the Fourier transform of $\rho(x)$ as follows
\begin{equation}
\tilde\rho(\omega)=\int^\infty_{-\infty}e^{-i\omega x}\rho(x)dx.
\end{equation}
\index{Fourier transform}
Using the formula $\int\pi^{-1}n/(x^2+n^2)\exp(-ix\omega)dx
=\exp(-n\vert\omega\vert)$, one can rewrite (\ref{eq:xxxfre}) as follows
$$\rho(\omega)(1+e^{-2\vert\omega\vert})=e^{-\vert\omega\vert}.$$
Thus $\tilde\rho(\omega)=1/(2\cosh\omega)$ and
\begin{equation}
\rho(x)={1\over 2\pi}\int^\infty_{-\infty}e^{i\omega x}
\tilde\rho(\omega)d\omega=
{1\over 4}{\rm sech}({\pi x\over 2}).
\end{equation}
Substituting this into (\ref{eq:xxxmagfre}) we have
\begin{equation}
e=-\vert J\vert (\ln 2-{1\over 4})=-0.443147\vert J\vert,\quad s_z=0.
\end{equation}
It can be shown 
that the case $B=\infty$ is the true ground state at
$h=0$ and that the magnetization is zero. Thus an analytical
result for a one-dimensional antiferromagnet is obtained (\cite{hulthen38}).
The first neighbor correlation is derived from this result
\begin{equation}
<S_i^zS_{i+1}^z>={1\over 12}(1-4\ln 2)=-0.14771573.
\end{equation}
An analytic expression for the second neighbor
correlation function is also known for this model (\cite{tak77}),
\begin{equation}
<S_i^zS_{i+2}^z>={1\over 12}(1-16\ln 2+9\zeta(3))=0.06067977.
\end{equation}
\index{second neighbor correlation}
\par\noindent
This was calculated from the ground state energy of the Hubbard model. \par
\noindent
Hulthen's solution is the ground state of XXX antiferromagnet at zero 
magnetic field. Griffiths calculated the magnetization curve of this model
(\cite{grif64}).  des Cloizeaux and Pearson calculated the elementary 
excitation away from this ground state (\cite{cloi62}).

\section{String solution of infinite system}
The elementary excitation for the $M=1$ case is
\begin{equation}
E-E_0=J(1-\cos K)+2h.
\end{equation}
This is the spin wave excitation of the ferromagnetic Heisenberg model.
Bethe found that the bound state of spin waves exists (Bethe, 1931),
\begin{equation}
x_j=\alpha+i(n+1-2j), j=1,2,...,n.
\end{equation}
\index{bound state}
The energy and momentum of this excitation are
\begin{equation}
E=E_0+{2nJ\over \alpha^2+n^2}+2nh,\quad K={1\over i}
\ln\Bigl({\alpha+in\over \alpha-in}\Bigr).
\end{equation}
Thus the dispersion relation is
\begin{equation}
E=E_0+{2J\over n}(1-\cos K)+2nh.
\label{eq:xxxeng}
\end{equation}
These string solutions play an 
essential role in the thermodynamics of soluble models.
\index{string solution}
The author obtained the thermodynamic Bethe ansatz equation for the XXX model
(\cite{tak71a}).
The wave function is
\begin{eqnarray}
&&f(n_1,n_2,...,n_M)=\sum_P A(P)
\prod_{j=1}^M\Bigl({x_{Pj}+i\over x_{Pj}-i}\Bigr)^{n_j},\nonumber\\
&&A(P)=D\epsilon(P)
\prod_{j<l}(x_{Pj}-x_{Pl}-2i).\label{eq:xxxwvf}
\end{eqnarray}
We consider the wave function in the equation (\ref{eq:xxxwvf}) assuming that
$N$ is infinity. Particle coordinates $n_j$ move from $-\infty$ to $\infty$.
Assume that $\Im k_1\ge\Im k_2\ge  ...\ge \Im k_n$. The wave function
is written as follows
\beqa
&&f(n_1,n_2,...,n_M)=(z_1z_2..z_M)^{n_1}\sum_PA(P)
\prod_{j=2}^{M}(\prod_{l=j}^M z_{Pl})^{n_{j+1}-n_j},\nonumber\\
&&A(P)=\epsilon(P)\prod_{j<l}(x_{Pj}-x_{Pl}-2i),~~
z_j=e^{ik_j}=({x_j+i\over x_j-i}).\label{eq:wfxxx}
\enqa
In the infinite system $f$ satisfies the following boundary 
conditions
\beqa
&&|\lim_{n_1,...,n_r\to -\infty} f(n_1,n_2,...,n_M)|<\infty,\nonum\\
&&|\lim_{n_{M-r+1},...,n_M\to \infty} f(n_1,n_2,...,n_M)|<\infty,
\enqa
From this condition we find that $|z_1z_2..z_M|=1$ and that 
$A(P)=0$ if one of $|\prod_{l=j}^M z_{Pl}|$ is greater than 1. 
From the normalizability condition of the wave function we have
\beq
A(I)\ne 0,~~ A(P\ne I)=0 ,\quad\vert z_1z_2...z_n\vert=1.
\enq
These conditions are satisfied only if
\beq
x_j=\alpha+(n+1-2j)i,\quad j=1,2,...,n.
\enq
From  $|z_1z_2..z_M|=|{\alpha+ni\over \alpha-ni}|=1$, we find that $\alpha$ 
must be real. The condition 
$$(\prod_{l=j+1}^M z_{l})=|{\alpha+(n-2j)i\over \alpha-ni}|\le 1, 
~~j=1,2,...,n-1$$ 
is automatically satisfied for general $n$. 
Thus strings with arbitrary length are possible for the XXX Heisenberg chain.
Moreover, the  following type of string is impossible
\beq
x_1=\alpha+(2+\beta)i,x_2=\alpha+\beta i,
x_3=\alpha -\beta i,x_4=\alpha-(2+\beta)i,
\enq
where $\alpha$,$\beta$ are real and $\beta\ne 1$.

\section{String hypothesis for a long XXX ring}
\label{sec:secXXX2}
We assume that all rapidities $x_j$ belong to bound states with $n=1,2,...$.
For a bound states of $n$-$x$'s the real parts of all $x$'s are the same and
imaginary parts are $(n-1)i,(n-3)i,...,-(n-1)i$ within the accuracy of
$O(\exp(-\delta N))$. This assumption seems to be too strong and there are
some counter examples in special cases. This is a very controversial point of
thermodynamic Bethe-ansatz equations for soluble models, except in the repulsive
Boson case, which has no the string solution. But equations obtained using 
the string hypothesis seem to give the correct free energy and other
thermodynamic quantities. \par
Consider the case where $M_n$ bound states of $n-x$'s exist. We designate $x$'s
as
\beq
x_\alpha^{n,j},\quad \alpha=1,2,...,M_n,$$
$$x_\alpha^{n,j}=x_\alpha^{n}+i(n+1-2j)+{\rm deviation}.
\enq
From (\ref{eq:xxxal2}),
$$e^N(x_\alpha^{n,j})=\prod_{(m,\beta)\ne(n,\alpha)}
e({x_\alpha^{n,j}-x_\beta^m\over m-1})
e({x_\alpha^{n,j}-x_\beta^m\over m+1})\prod_{j'\ne j}
e({x_\alpha^{n,j}-x_\alpha^{n,j'}\over 2}),$$
\beq j=1,2,...,n.\enq
Here $e(x)\equiv (x+i)/(x-i)$. The last product is delicate because the
numerator or denominator may become very small. If we take the 
product of these
$n$ equations, these delicate terms are canceled and we have
\beq
e^N(x^n_\alpha/n)=\prod_{j=1}^n e^N(x_\alpha^{n,j})=\prod_{(m,\beta)\ne
(n,\alpha)}E_{nm}(x_\alpha^n-x_\beta^m),
\label{eq:xxxstring}\enq
where
\beq
E_{nm}(x)\equiv\left\{\matrix{
e({x\over \vert n-m\vert})e^2({x\over \vert n-m\vert+2})
e^2({x\over \vert n-m\vert+4})...e^2({x\over n+m-2})e({x\over n+m})\cr
{\rm for}\quad n\ne m,\cr
e^2({x\over 2})e^2({x\over 4})...e^2({x\over 2n-2})e({x\over 2n})\quad{\rm for}
\quad n=m.}\right.\label{eq:funcE}
\enq
The logarithm of these equations gives
\beq
N\theta(x_\alpha^n/n)=2\pi I_\alpha^n+\sum_{(m,\beta)\ne
(n,\alpha)}\Theta_{nm}(x_\alpha^n-x_\beta^m),
\enq
where
\beq
\theta(x)\equiv 2\tan^{-1}(x),
\enq
and
\beq
\Theta_{nm}(x)\equiv\left\{\matrix{\theta({x\over \vert n-m\vert})
+2\theta({x\over \vert n-m \vert +2})+...+2\theta({x\over n+m-2})
+\theta({x\over n+m})\cr
{\rm for}\quad n\ne m,\cr
2\theta({x\over 2})+2\theta({x\over 4})+...+2\theta({x\over 2n-2})
+\theta({x\over 2n})\quad {\rm for}\quad n=m.}\right.
\label{eq:funcTheta}
\enq
$I_\alpha^n$ is an integer (half-odd integer) if $N-M_n$ is odd (even) and
should satisfy
\beq
\vert I_\alpha^n\vert\le{1\over 2}(N-1-\sum_{m=1}^\infty t_{nm}M_m), \quad
t_{nm}\equiv 2Min(n,m)-\delta_{nm}.
\label{eq:xtint}
\enq
We can prove the number of sets $\{I_\alpha^n\}$ is
$C^N_M-C^N_{M-1}$, under the condition $M=
\sum_{n=1}^\infty nM_n$. Here $C^N_M$ is the binomial coefficient 
defined by $N!/(M!(N-M)!)$. 
For details, see Appendix A of (\cite{tak71a}). 
The energy of this state is given by
\beq
E(\{I_\alpha^n\})=N(-h-{J\over 4})+\sum_{n,\alpha}
({2Jn\over (x^n_\alpha)^2+n^2}+2hn).
\enq
We can construct wave functions through (\ref{eq:xxxfunc}), 
(\ref{eq:xxxf2}) and (\ref{eq:xxxA}) at $S=S_z=N/2-M$.
The wave functions for $S_z=S-1,S-2,...,
-S$ are obtained by applying the 
operator $S^-_{\rm total}$. The energy for these cases are
$E-2h,E-4h,...,E-2(N-2M)h$. Then the total number of states which are
generated by the string assumption and the descending operator at $S_z=N/2-M$ is
$C^N_M$. Therefore the total number of states is $2^N$. This coincides
with the true total number of states. It is expected that all eigen
functions constructed in the above way should be a complete set.
The partition function of this system is written as follows
\beq
{\cal Z}=\sum_{M=0}^{[N/2]}{1-\exp(-2(N+1-2M)h/T)\over 1-\exp(-2h/T)}
\sum_{\{I_\alpha^n\}}\exp[-T^{-1}E(\{I_\alpha^n\})].
\enq
The free energy is given by $G=-T\ln{\cal Z}$. We define functions $h_n(x)$ by
\beq
h_n(x)\equiv \theta_n(x)-N^{-1}\sum_{(m,\alpha)}\Theta_{nm}(x-x^m_\alpha).
\label{eq:xt35}
\enq
We can define holes in $n$-string sea by the solution of
\beq
2\pi J_\beta^n/N=h_n(x_\beta^n),
\enq
where $J_\beta^n$ are omitted integers or half-odd integers in the region.

\section{Thermodynamic Bethe-ansatz equations for XXX chain}
In the thermodynamic limit, define distribution functions of $n$-strings
and holes of $n$-string as $\rho_n(x)$ and $\rho_n^h(x)$. The number of strings
and holes between $x$ and $x+dx$ is $\rho_n(x)Ndx$ and  $\rho_n^h(x)Ndx$, 
respectively.
Thus we have
\beq
2\pi\int^x \rho_n(t)+\rho^h_n(t)dt=\theta_n(x)-\sum_{m=1}^\infty
\int^\infty_{-\infty}
\Theta_{nm}(x-y)\rho_m(y)dy.
\enq
Differentiating by $x$, we obtain the integral equation
\beq
a_n(x)=\rho_n(x)+\rho^h_n(x)+\sum_m\int^\infty_{-\infty} 
T_{nm}(x-y)\rho_m(y)dy,
\label{eq:xtequ38}
\enq
where $T_{nm}(x)$ is a function defined by
\beq
T_{nm}(x)\equiv \left\{\matrix{a_{|n-m|}(x)+2a_{|n-m|+2}(x)
+2a_{|n-m|+4}(x)+...\cr
+2a_{n+m-2}(x)+a_{n+m}(x)~~~~{\rm for}\quad n\ne m,\cr 
2a_{2}(x)+2a_{4}(x)+...
+2a_{2n-2}(x)+a_{2n}(x)\quad {\rm for}\quad n=m.}\right.
\label{eq:xtequ39}
\enq
$a_n(x)$ is a function defined by
\beq
a_n(x)\equiv{1\over \pi}{n\over x^2+n^2},~~a_0(x)\equiv\delta(x).
\enq
The energy per site is
\beqa
&&e=-({J\over 4}+h)+\sum_{n=1}^\infty\int^\infty_{-\infty} g_n(x)\rho_n(x)dx,\nonum\\
&&g_n(x)\equiv 2\pi J a_n(x)+2nh.
\label{eq:xtequ41}
\enqa
The total entropy per site is
\beq
{\it s}=\sum^\infty_{n=1}\int^\infty_{-\infty} 
\rho_n(x)\ln(1+{\rho_n^h(x)\over\rho_n(x)})
+\rho_n^h(x)\ln(1+{\rho_n(x)\over\rho_n^h(x)})dx.
\enq
$e-T{\it s}$ should be minimized at the thermodynamic equilibrium.
Consider the functional variation of the free energy with respect to 
$\rho_n(k)$ and $\rho_n^h(k)$,
\beqa
&&0=\delta e-T\delta {\it s}=\sum_{n=1}^\infty\int dx\nonum\\
&&[g_n(x)-T\ln(1+{\rho_n^h(x)\over \rho_n(x)})]\delta\rho_n(x)
-T\ln(1+{\rho_n(x)\over \rho_n^h(x)})\delta\rho_n^h(x).\nonum\\
\label{eq:xt43}
\enqa
From equation (\ref{eq:xt35}) we have
\beq
\delta\rho_n^h(x)=-\delta\rho_n(x)-\sum_m\int T_{nm}(x-y)\delta\rho_m(y)dy.
\enq
Substituting these into (\ref{eq:xt43}) yields
\beq
0=T\sum_{n=1}^\infty\int\Bigl\{{g_n(x)\over T}-\ln\eta_n(x)+\sum_{m=1}\int
T_{nm}(x-y)\ln(1+\eta_m^{-1}(y))dy\Bigr\}\delta\rho_n(x)dx,
\enq
where $\eta_n(x)\equiv\rho_n^h(x)/\rho_n(x)$. Thus we have integral
equations for an infinite number of unknown $\eta_n(x)$
\beq
\ln\eta_n(x)={g_n(x)\over T}+\sum^\infty_{m=1}T_{nm}*\ln(1+\eta_m^{-1}(x)).
\label{eq:xtequ46}
\enq
In the theory of Bethe ansatz equations we encounter very frequently 
the integration of the following type
$$\int^\infty_{-\infty} a(x-y)b(y)dy.$$
This is a convolution of two functions $a(x)$ and $b(x)$. 
In the above and hereafter 
we write the convolution as $a*b(x)$,
\beq
 a*b(x)\equiv \int^\infty_{-\infty} a(x-y)b(y)dy.
\label{eq:conv1}
\enq
The free energy per site becomes as follows
$$f=e-T{\it s}=-({J\over 4}+h)+\sum_{n=1}^\infty\int 
g_n\rho_n-T[\rho_n\ln(1+\eta_n)
+\rho_n^h\ln(1+\eta_n^{-1})]dx.$$
We eliminate $\rho_n^h$ using (\ref{eq:xtequ38})
\beqa
&&f=-({J\over 4}+h)-T\sum_{n=1}^\infty\int\ln(1+\eta_n^{-1})
a_n(x)\nonum\\
&&+\rho_n
[\ln\eta_n-{g_n\over T}-T_{nm}*\ln(1+\eta_m^{-1})]dx.
\label{eq:xxxfen1}
\enqa
Using (\ref{eq:xtequ46}), one sees the inside of the bracket on r.h.s. 
is zero. So we have
\beq
f=-({J\over 4}+h)-T\sum_{n=1}^\infty\int
a_n(x)
\ln(1+\eta_n^{-1}(x))dx.
\label{eq:xtfe}
\enq
From the $n=1$ case of (\ref{eq:xtequ46}) we have
\beq
\ln(1+\eta_1)={2\pi Ja_1(x)+2h\over T}+\sum_{l=1}^\infty(a_{l-1}+a_{l+1})
*\ln(1+\eta^{-1}_l).
\enq
Operating $\int dx s(x)$ on this equation yields
$$\int dx s(x)\ln(1+\eta_1)={2\pi J\over T}\int s(x)a_1(x)dx+{h\over T}$$
$$+\sum_{l=1}^\infty\int a_l(x)\ln(1+\eta^{-1}_l(x))dx.$$
Then equation (\ref{eq:xtfe}) is transformed as follows
\beq
f=J(\ln2-{1\over 4})-T\int s(x)\ln(1+\eta_1(x))dx.
\label{eq:xtfe2}
\enq

Solutions $\eta_n$ of (\ref{eq:xtequ46}) are functions of
$x$, $J$, $T$ and $h$.
Differentiating (\ref{eq:xtequ46}) with respect to $J$ yields
\beq
{2\pi J a_n(x)\over T}=
{1\over \eta_n}{\partial \eta_n\over \partial J}+\sum_mT_{nm}*
{1\over (1+\eta_m)\eta_m}{\partial \eta_m\over \partial J}.
\enq
Comparing this with (\ref{eq:xtequ38}) we have
\beq
\rho_n={T\over 2\pi}{1\over (1+\eta_n)\eta_n}
{\partial \eta_n\over \partial J},\quad
\rho_n+\rho_n^h={T\over 2\pi}{\partial \ln\eta_n\over \partial J}.
\enq
By the definition (\ref{eq:xtequ39}),
\beq
a_{1}*(T_{n-1,m}+T_{n+1,m})-(a_{0}+a_{2})*T_{n,m}=
(\delta_{n-1,m}+\delta_{n+1,m})a_{1}.
\label{eq:xt53}
\enq
Using equations (\ref{eq:xtequ46}), (\ref{eq:xt53}) yields
\beq
(a_{0}+a_{2})*\ln \eta_1(x)={2\pi J a_1(x)\over T}
+a_{1}*\ln(1+\eta_2(x)),
\label{eq:xt54a}
\enq
\beq
(a_{0}+a_{2})*\ln \eta_n(x)=a_{1}*\ln(1+\eta_{n-1}(x))(1+\eta_{n+1}(x)),\quad
n=2,3,....
\label{eq:xt54b}
\enq
Equations (\ref{eq:xt54a}) and(\ref{eq:xt54b}) are not complete to determine
all of the $\eta_n(x)$, as 
they do not contain $h$.  Take the $n=1$ case of (\ref{eq:xtequ46})
\beq
\ln\eta_1={2\pi J a_1(x)+2h\over T}+
a_{2}*\ln(1+\eta^{-1}_1)+\sum_{j=2}^\infty(a_{j-1}+a_{j+1})*\ln(1+\eta_j^{-1}).
\enq
Substituting (\ref{eq:xt54a}), (\ref{eq:xt54b})
we can eliminate $\eta_j, j<n$ for a given integer $n$,
\beqa
&&{2h\over T}=a_{n}*\ln\eta_{n+1}-a_{n+1}*\ln(1+\eta_n)-a_{n+2}*
\ln(1+\eta_{n+1}^{-1})\nonum\\
&&-\sum_{l=n+2}^\infty(a_{l-1}+a_{l+1})*\ln(1+\eta^{-1}_l).
\enqa
Thus we have
\beqa
&&\ln\eta_{n+1}={2h\over T}+a_{1}*\ln\eta_n+a_{2}*\ln(1+\eta_{n+1}^{-1})
\nonum\\
&&+\sum_{l=n+2}^\infty(a_{l-n-1}+a_{l-n+1})*\ln(1+\eta^{-1}_l).
\enqa
For large $n$, $\ln(1+\eta^{-1}_n)\simeq o(n^{-2})$ and therefore:
\beq
\lim_{n\to \infty}\ln\eta_{n+1}-a_{1}*\ln\eta_n={2h\over T},
\enq
or
\beq
\lim_{n\to \infty}{\ln\eta_n\over n}={2h\over T}.
\enq
Thus the following equations determine $\eta_n$,
\beqa
&&\ln\eta_1(x)={2\pi J\over T}s(x)+s*\ln(1+\eta_2(x)),
\label{eq:xt59a}\\
&&\ln\eta_n(x)=s*\ln(1+\eta_{n-1}(x))(1+\eta_{n+1}(x)),
\label{eq:xt59b}\\
&&\lim_{n\to \infty}{\ln\eta_n\over n}={2h\over T},
\label{eq:xt59c}
\enqa
where
\beq
s(x)={1\over 4}{\rm sech}({\pi x\over 2}).
\label{eq:xt59d}
\enq

\section{Some special cases and expansions}
\subsection{$J/T\to {\it 0}$ case}
In the limit $J/T\to 0$ and $h/T\ge 0$ we can expect that $\eta_n(x)$ is
independent of $x$, because there are no $x$ dependent terms in equations
(\ref{eq:xt59a}), (\ref{eq:xt59b} and (\ref{eq:xt59c}).
As $\int dxs(x)=1/2$, equations
(\ref{eq:xt59a},\ref{eq:xt59b},\ref{eq:xt59c}) become
\beqa
&&\eta_n^2=(1+\eta_{n-1})(1+\eta_{n+1}),
\label{eq:xt60a}\\
&&\eta_1^2=1+\eta_2,
\lim_{n\to \infty}\ln \eta_n/n={2h/T}.
\label{eq:xt60b}
\enqa
Equation (\ref{eq:xt60a}) is a difference equation of second order.
\index{difference equation}
It is similar
to a differential equation of the second order and contains two arbitrary
parameters. The general solution of this equation is
\beq
\eta_n=({az^n-a^{-1}z^{-n}\over z-z^{-1}})^2-1.
\enq
Parameters $a$ and $z$ are determined by (\ref{eq:xt60b}) and we have
$a=z,z=\exp(h/T)$ and 
\beq
\eta_n=({\sinh[(n+1)h/T]\over \sinh[h/T]})^2-1\quad {\rm for}\quad h>0$$
$$\eta_n=(n+1)^2-1\quad{\rm for}\quad h=0.
\enq
Substituting this into (\ref{eq:xtfe2}) we obtain the free energy,
magnetization and entropy
\beqa
&&f=-T\ln[2\cosh h/T], \quad m=2s_z=-\partial f/\partial h=\tanh h/T,\nonum\\
&&{\it s}=-\partial f/\partial T=\ln[2\cosh(h/T)]-(h/T)\tanh(h/T).
\label{eq:xt63}
\enqa
At $h=0$ the entropy per site is $\ln 2$. This corresponds to 
the fact that the
number of states per site is two.

\subsection{High temperature expansion or small $J$ expansion}
\index{high temperature expansion}
For the XXX chain, we can perform the high temperature expansion of the
free energy density from the definition
\beq
f/T=-N^{-1}\ln{\rm Tr}\exp(-{\cal H}/T).
\enq
This is expanded as a power series of $1/T$. Assume that ${\cal H}
={\cal H}_0+{\cal H}_1$ where ${\cal H}_0$ and ${\cal H}_1$ commute each other.
Then the exponential operator of ${\cal H}$ can be expanded as follows
\beq
\exp(-{\cal H}/T)=\exp(-{\cal H}_0/T)\Bigl(1-T^{-1}{{\cal H}_1\over 1!}
+T^{-2}{{\cal H}_2\over 2!}-+...\Bigr ).
\enq
Thus,
\beq
f/T=-N^{-1}\ln{\rm Tr}\exp(-{\cal H}_0/T)+{<{\cal H}_1>\over NT}
-{<{\cal H}^2_1>-<{\cal H}_1>^2\over 2!NT^2}$$
$$+{<{\cal H}^3_1>-3<{\cal H}_1>^2<{\cal H}_1>+2<{\cal H}_1>^3\over 3!NT^3}
-+...,
\enq
where
\beq
<X>\equiv {{\rm Tr}\exp(-{\cal H}_0/T)X\over {\rm Tr}\exp(-{\cal H}_0/T)}.
\enq
In the beginning of 1930's 
only the ferromagnetic model was
considered.
If we set
\beq
{\cal H}_0=-2h\sum_{l=1}S^z_l,\quad
{\cal H}_1=-J\sum_{l=1}^NS_l^xS_{l+1}^x+S_l^yS_{l+1}^y+S_l^zS_{l+1}^z,
\enq
for the Hamiltonian (\ref{eq:xxxham}),
we obtain the $J/T$ expansion of free energy at fixed $h/T$
\beqa
&&f/T=-\ln(2\cosh h/T)-{J\over 4T}\tanh^2(h/T)\nonum\\
&&-{J^2\over 32 T^2}(3+2\tanh^2(h/T)-3\tanh^4(h/T))+O((J/T)^3).
\label{eq:xt69}
\enqa
The calculation to higher orders can be done by the use of linked cluster
expansion. \index{linked cluster expansion}
Higher order terms are polynomials of $\tanh(h/T)$.
\par
Apparently the expression of the free energy in (\ref{eq:xt63})
coincides with the first
term of the above expansion. 
Writing $\ln(\eta_n+1)$ as the expansion
\beqa
&&\ln(\eta_n(x)+1)=\ln[{1\over \alpha_n-1}]
+\sum_{l=1}^\infty f_n^{(l)}[{J\over T}]^l,\nonum\\
&&\alpha_n\equiv {\sinh^2(h(n+1)/T)\over\sinh(hn/T)\sinh(h(n+2)/T)},
\enqa
we obtain an expansion of $\ln\eta_n(x)$
\beqa
&&\ln\eta_n(x)=\ln {\alpha_n\over \alpha_n-1}+[{J\over T}]\alpha_nf_n^{(1)}
\nonum\\
&&+[{J\over T}]^2(\alpha_nf_n^{(2)}+(\alpha_n-\alpha_n^2){(f_n^{(1)})^2\over 2})
+O([{J\over T}]^3).
\enqa
Substituting these expansions into
(\ref{eq:xt59a},\ref{eq:xt59b},\ref{eq:xt59c},\ref{eq:xt59d})
and taking first order terms in
$J/T$, linear integral equations for $ f_n^{(1)}(x)$ are obtained
\beqa
&&\alpha_1 f_1^{(1)}(x)-s* f_2^{(1)}(x)=2\pi s(x),
\label{eq:xt72a}\\
&&\alpha_n f_n^{(1)}(x)-s*(f_{n-1}^{(1)}(x)+f_{n+1}^{(1)}(x))=0,
\label{eq:xt72b}\\
&&\lim_{n\to\infty}{\alpha_nf_n(x)\over n}=0.
\label{eq:xt72c}
\enqa
The r.h.s. of these equations are inhomogeneous terms of the integral 
equations.
The Fourier transform of these equations are
\beq
(e^{\vert\omega\vert}+e^{-\vert\omega\vert})\alpha_n\tilde f_n^{(1)}
(\omega)=  \tilde f_{n-1}^{(1)}(\omega)+\tilde f_{n+1}^{(1)}(\omega).
\enq
The general solution of this difference equation is
\beqa
&&\tilde f_n^{(1)}(\omega)=\nonum\\
&&A(\omega)[{\sinh((n+2)h/T)\over\sinh((n+1)h/T)}
e^{-n\vert\omega\vert}
-{\sinh(nh/T)\over\sinh((n+1)h/T)}e^{-(n+2)\vert\omega\vert}]\nonum\\
&&+B(\omega)[{\sinh((n+2)h/T)\over\sinh((n+1)h/T)}e^{n\vert\omega\vert}
-{\sinh(nh/T)\over \sinh((n+1)h/T)}e^{(n+2)\vert\omega\vert}].\nonum\\
\enqa
From the boundary conditions we have
\beq
A(\omega)={\pi\over \cosh h/T},\quad B(\omega)=0.
\enq
Thus
\beqa
&&\tilde f_1^{(1)}(\omega)={\pi\over \cosh(h/T)}[{\sinh 3h/T\over \sinh 2h/T}
e^{-\vert\omega\vert}-{\sinh h/T\over \sinh 2h/T}
e^{-3\vert\omega\vert}],\nonum\\
&&f_1^{(1)}(x)={\pi\over \cosh(h/T)}[{\sinh 3h/T\over \sinh 2h/T}
a_1(x)-{\sinh h/T\over \sinh 2h/T}a_3(x)].
\enqa
Substituting this into (\ref{eq:xtfe2}) we obtain the second term
of the $J/T$ expansion (\ref{eq:xt69}).
The higher order terms can be calculated by solving the linear
integral equations for $f_n^{(2)}, f_n^{(3)},...$. The equations are 
similar to (\ref{eq:xt72a},\ref{eq:xt72b},\ref{eq:xt72c})
except for the inhomogeneous terms, which are given by lower order
$f_n^{(l)}$.

\subsection{Low temperature limit}
At low temperature, $\ln \eta_n$ diverges as $1/T$. So we should define
the following functions
\beq
\epsilon_n(x)=T\ln \eta_n(x).
\enq
The integral equations become
\beqa
&&\epsilon_1(x)=2\pi J s(x)+s*T\ln(1+\exp({\epsilon_2(x)\over T})),\nonum\\
&&\epsilon_n(x)=s*T\ln(1+\exp({\epsilon_{n-1}(x)\over T}))
(1+\exp({\epsilon_{n+1}(x)\over T})),\nonum\\
&&\lim_{n\to\infty}{\epsilon_n(x)\over n}=2h.
\enqa
The free energy expression becomes
\beqa
f&=&-({J\over 4}+h)-T\sum_{n=1}^\infty\int a_n(x)
\ln(1+\exp(-\epsilon_n(x)/T)dx\nonum\\
&=&J(\ln2-{1\over 4})-T\int s(x)\ln(1+\exp(\epsilon_1(x)/T))dx.
\enqa
The $T=0$ limit of these equations is
\beqa
&&\epsilon_1(x)=2\pi J s(x)+s*\epsilon_2^+(x),\nonum\\
&&\epsilon_n(x)=s*(\epsilon_{n-1}^+(x)+\epsilon_{n+1}^+(x)),\nonum\\
&&\lim_{n\to\infty}{\epsilon_n(x)\over n}=2h,
\enqa
\beqa
&&f=-({J\over 4}+h)+\sum_{n=1}^\infty\int a_n(x)
\epsilon_n^-(x)dx
=J(\ln2-{1\over 4})-\int s(x)\epsilon_1^+(x)dx,\nonum\\
&&\epsilon_n^+(x)\equiv \left\{\begin{array}{@{\,}ll}
\epsilon_n(x),& \mbox{ for  $\epsilon_n(x)\ge 0,$}\\
0,& \mbox{  for  $\epsilon_n(x)<0,$}. \end{array} \right.\nonum\\
&&\epsilon_n^-(x)\equiv \left\{\begin{array}{@{\,}ll}
0,& \mbox{ for  $\epsilon_n(x)\ge 0,$}\\
\epsilon_n(x),& \mbox{  for  $\epsilon_n(x)<0$}. \end{array} \right.
\enqa
\par
In the ferromagnetic case $J>0$ we have
\beq
\epsilon_n(x)=\epsilon_n^+(x)=2\pi J a_n(x)+2hn,\quad n=1,2,...,
\enq
and therefore $f=-({J\over 4}+h)$. This is the ground state energy of the
ferromagnetic case.  \par
In the antiferromagnetic case $J<0$ we have
\beq
\epsilon_n(x)=\epsilon_n^+(x)=
a_{n-1}*\epsilon_1^+(x)+2(n-1)h,\quad n=2,3,....
\enq
The equation which determines $\epsilon_1$ is
\beq
\epsilon_1(x)=-2\pi\vert J\vert s(x)+h
+\int_{\vert y\vert>B}R(x-y)\epsilon_1(y)dy,\quad \epsilon_1(\pm B)=0.
\enq
In the limit of $h\to 0$ $B$ becomes infinite. We have
$\epsilon_1(x)=-2\pi\vert J\vert s(x)$ and $f=J(\ln2-{1\over 4})$.

\subsection{Fugacity expansion}
\index{fugacity expansion}
In the case of very large $h$ the free energy can be 
expanded as a power series of
$z=\exp(-h/T)$, $z$ is the called the fugacity. From equation (\ref{eq:xtequ41})
and (\ref{eq:xtequ46}) we have expansions of
$\eta_n^{-1}$ as follows
\beq
\eta_n^{-1}=z^{2n}\exp(-{2\pi J\over T}a_n(x))
\exp[-T_{nm}*(\eta_m^{-1}-{1\over 2}\eta_m^{-2}+-...)].
\enq
The expansion of $\eta_1^{-1}$ and $\eta_2^{-1}$ up to $z^4$ is
\beqa
&&\eta_1^{-1}=z^2\exp(-{2\pi J\over T} a_1(x))
(1-
z^2\int a_2(x-y)\exp(-{2\pi J\over T}a_1(y))dy)+O(z^6),\nonum\\
&&\eta_2^{-1}=z^4\exp(-{2\pi J\over T}a_2(x))+O(z^6).
\label{eq:xt88}
\enqa
As $\eta_n^{-1}$ becomes small, (\ref{eq:xtfe}) is more convenient than
(\ref{eq:xtfe2})
$$f=-{J\over 4}-h-T\sum_{n=1}^\infty\int dx a_n(x)
(\eta_n^{-1}-{1\over 2}\eta_n^{-2}+-...). $$
Substituting (\ref{eq:xt88}), we obtain
\beqa
&&f=-{J\over 4}-h-z^2T\int a_1(x)\exp(-{2\pi J\over T}a_1(x))dx
\nonum\\
&&-z^4T\Bigl\{\int a_2(x)\exp(-{2\pi J\over T}a_2(x))
-{1\over 2}a_1(x)\exp(-{4\pi J\over T}a_1(x))dx\nonum\\
&&-\int dx\int dy
a_1(x)\exp\Bigl[-{2\pi J\over T}(a_1(x)+a_1(y))\Bigr]a_2(x-y)\Bigr\}+O(z^6).
\nonum\\
\enqa
Putting $x=\tan u, y=\tan v$,
\beqa
&&(f+{J\over 4}+h)/T\nonum\\
&&=z^2e^{-K}I_0(K)+z^4\Bigl\{-{1\over 2}e^{-2K}I_0(2K)
+e^{-K/2}I_0(K/2)\nonum\\
&&-{1\over\pi^2}\int^\pi_0\int^\pi_0 {e^{2K(1-\cos \omega_1\cos \omega_2)}
(1-\cos \omega_1\cos \omega_2)\over 1-2\cos \omega_1\cos \omega_2+
\cos^2\omega_1}d\omega_1d\omega_2\Bigr\}\nonum\\
&&+O(z^6),
\enqa
where $K\equiv J/T, \omega_1=\pi+u+v,\omega_2=u-v$ and
$I_0(x)$ is modified Bessel function. This result is the same as that of
Katsura (\cite{katura65}).
\par
The strings are stable in a chain of infinite length in the 
case of very few down spins. Some counter examples of string 
assumption are found in some special limit. Nevertheless the thermodynamic 
Bethe ansatz equation seems to give the exact free energy in the case 
where the density of down-spins is comparable to that of up-spins. 
This non-linear integral 
equation contains an infinite number of unknown functions. 
To solve this equation 
one needs to do numerical calculations by computer.


\chapter{Thermodynamics of the XXZ model}

\section{Symmetry of the Hamiltonian}
\index{XXZ model}
In this section we consider the following Hamiltonian
\beqa
&&{\cal H}(J,\Delta,h)
=-J\sum_{l=1}^NS_l^xS_{l+1}^x+S_l^yS_{l+1}^y+\Delta S_l^zS_{l+1}^z
-2h\sum_{l=1}S^z_l,\nonum\\
&&h\ge 0,~~{\bf S}_{N+1}\equiv {\bf S}_1.
\label{eq:xxzham}
\enqa
This Hamiltonian contains an additional parameter $\Delta$. 
The case $\Delta=0$ is called the XY model, which can be mapped to 
non-interacting fermions making it possible to calculate many physical quantities (\cite{lsm61}, \cite{katura62}). 
The case $\Delta=1$ is the XXX
model and was treated in the previous chapter. 
The limit of very large $\Delta$ is the Ising model.
\index{Ising model}
The generalization of Bethe's method to $\Delta\ne  1$ was done by
Orbach and Walker (\cite{orb58} \cite{walk59}).
Yang and Yang investigated the ground state of this
model in detail (\cite{yang66}).
Bonner and Fisher investigated this
model using the diagonalization method up to $N=12$
(\cite{B&F64}).
In this Hamiltonian the magnetic field is
applied in the z-direction. For a magnetic field in a different
direction, the exact solution is not known.  \par
Let us consider the following unitary transformation:
\begin{equation}
{\cal H}(J,\Delta,h)=U_1{\cal H}(J,\Delta,-h)U_1^{-1},
\quad U_1\equiv\prod_{l=1}^\infty 2S_l^x=U_1^{-1}.
\end{equation}
By this unitary transformation $S_{total}^z$ changes its sign and we can treat
the $N\ge M>N/2$ case.
In the case of even $N$ we can show that
\begin{equation}
{\cal H}(-J,-\Delta,h)=U_2{\cal H}(J,\Delta,h)U_2^{-1},
\quad U_2\equiv\prod_{l=even} 2S_l^z=U_2^{-1}.\label{eq:u2}
\end{equation}
By this unitary transformation $S_l^x,S_l^y,S_l^z$ change to
$-S_l^x,-S_l^y,S_l^z$ at $l=even$.

\section{Bethe ansatz wave function}
Consider the state where all spins are up and the total
$S^z$ is $N/2$
\begin{equation}
{\cal H}\vert 0>=E_0\vert 0>,\quad E_0=-J\Delta N/4-Nh.
\end{equation}
Writing a general state $|\Psi>$ in terms of a wave function $f$, as in 
equation (\ref{eq:xxxfunc}), the eigenvalue condition can be expressed as
\beqa
&&0=-{J\over 2}\sum_j(1-\delta_{n_j+1, n_{j+1}})
\Bigl\{f(n_1,.,n_j+1,n_{j+1},.,n_M)\nonum\\
&&+f(n_1,.,n_j,n_{j+1}-1,.,n_m)\Bigr\}\nonum\\
&&+\Bigl\{E_0-E+(J\Delta+2h)M-J\Delta\sum_j\delta_{n_j+1,n_{j+1}} \Bigr\}
f(n_1,n_2,.,n_M).\nonum\\
\enqa
For a wave function of the type eq. (\ref{eq:xxxf2}) to be the eigenstate, 
set $E$ and $A(P)$ to be
\beqa
&&E=E_0+\sum_{j=1}^M[J(\Delta-\cos k_j)+h],\label{eq:xxzene}\\
&&0=A(P)(e^{ik_{Pj}}+e^{-k_{P(j+1)}}-2\Delta)e^{k_{P(j+1)}}\nonum\\
&&+A(P(j,j+1))(e^{ik_{P(j+1)}}+e^{-k_{Pj}}-2\Delta)e^{k_{Pj}},\\
&&A(P)=\epsilon(P)\prod_{l<j}(e^{i(k_{Pl}+k_{Pj})}+1-2\Delta e^{ik_{Pl}}).
\enqa
The periodic boundary condition is as follows,
\beqa
\exp(ik_{j}N)=(-1)^{M-1}\prod_{l\ne j}{\exp[i(k_j+k_l)]+1-2\Delta\exp(ik_j)
\over \exp[i(k_j+k_l)]+1-2\Delta\exp(ik_l)},\nonum\\
\quad j=1,2,...,M.
\label{eq:xxzarg}
\enqa
This is a set of complicated coupled equations for $M$ unknowns. If we have
a solution of this set, we have one eigenstate and its energy eigenvalue
and total momentum.  If we set rapidity parameters $x_j$ as $\cot(k_j/2)$
\begin{equation}
\exp(ik_j)=\Bigl({x_j+i\over x_j-i}\Bigr),
\label{eq:xxxtra}
\end{equation}
the phase factor
$${\exp[i(k_j+k_l)]+1-2\Delta\exp(ik_j)
\over \exp[i(k_j+k_l)]+1-2\Delta\exp(ik_l)}$$
cannot be written as a function of $x_j-x_l$ except in the case 
$\Delta=1$. This is not convenient. 
If we set
\begin{equation}
\exp(ik_j)={\sin{\phi\over 2}(x_j+i)\over
\sin{\phi\over 2}(x_j-i)},
\label{eq:xxztra}
\end{equation}
in place of (\ref{eq:xxxtra}), the phase factor becomes
$${\cos{\phi\over 2}(x_j+x_l)(\cosh \phi-\Delta)+
(\Delta\cos{\phi\over 2}(x_j-x_l+2i)-\cos{\phi\over 2}(x_j-x_l))\over
\cos{\phi\over 2}(x_j+x_l)(\cosh \phi-\Delta)+
(\Delta\cos{\phi\over 2}(x_l-x_j+2i)-\cos{\phi\over 2}(x_l-x_j))}.$$
At $\cosh \phi-\Delta=0$, this phase factor becomes a function
of $x_j-x_l$ and independent of $x_j+x_l$. It is written as
$$\sin{\phi\over 2}(x_j-x_l+2i)/\sin{\phi\over 2}(x_j-x_l-2i).$$
Then for $\Delta>1$ (\ref{eq:xxzarg}) becomes
\beqa
\Bigl({\sin{\phi\over 2}(x_j+i)\over \sin{\phi\over 2}(x_j-i)}\Bigr)^N
=\prod_{l\ne j}
{\sin{\phi\over 2}(x_j-x_l+2i)\over\sin{\phi\over 2}(x_j-x_l-2i)},\nonum\\
\phi=\cosh^{-1}\Delta,\quad \phi>0.
\label{eq:xxzbethe}\enqa
For $\Delta<-1$ we set
\begin{equation}
\exp(ik_j)=-{\sin{\phi\over 2}(x_j+i)\over
\sin{\phi\over 2}(x_j-i)}.
\end{equation}
Equation (\ref{eq:xxzarg}) becomes
\beqa
\Bigl({\sin{\phi\over 2}(x_j+i)\over \sin{\phi\over 2}(x_j-i)}\Bigr)^N
=\prod_{l\ne j}
{\sin{\phi\over 2}(x_j-x_l+2i)\over\sin{\phi\over 2}(x_j-x_l-2i)},\nonum\\
\phi=\cosh^{-1}(-\Delta),\quad \phi>0.
\label{eq:xxzarg2}
\enqa
In the case $1>\Delta>-1$ we set
\begin{equation}
\exp(ik_j)=-{\sinh{\gamma\over 2}(x_j+i)\over
\sinh{\gamma\over 2}(x_j-i)},
\end{equation}
yielding the Bethe ansatz equations
\beqa
&&\Bigl({\sinh{\gamma\over 2}(x_j+i)\over
\sinh{\gamma\over 2}(x_j-i)}\Bigr)^N
=\prod_{l\ne j}
{\sinh{\gamma\over 2}(x_j-x_l+2i)\over\sinh{\gamma\over 2}(x_j-x_l-2i)},
\nonum\\
&&\gamma=\cos^{-1}(-\Delta),~~\pi>\gamma>0.\label{eq:xxzarg3}
\enqa

\section{String solutions at $\Delta>{\it 1}$}
\index{string solution} 
By the transformation (\ref{eq:xxztra}) the wave function and eigenvalue are 
written as follows,
\beqa
&&f(n_1,n_2,...,n_M)\nonum\\
&&=\sum_P \epsilon(P)
\prod_{j<l}\sin{\phi\over 2}(x_{Pj}-x_{Pl}+2i)
\prod_{j=1}^M({\sin{\phi\over 2}(x_{Pj}+i)\over
\sin{\phi\over 2}(x_{Pj}-i)})^{n_j},
\enqa
\begin{equation}
E=E_0+\sum_{j=1}^M(2h+{J\sinh^2 \phi\over \cosh\phi-\cos \phi x_j}),
~~K=\sum_j2\cot^{-1}{\tan(\phi x_j/2)\over \tanh(\phi/2)}.
\end{equation}
The following string solutions are possible for complex $x_j$'s for the 
$N=\infty$ case from the normalizability of the wave function,
\begin{equation}
x_j=\alpha+(M+1-2j)i,
\end{equation}
where $\alpha$ is a real number at $-Q<\alpha\le Q,~Q\equiv \pi/\phi$.
The total momentum and energy is given by
\begin{equation}
K=2\cot^{-1}{\tan(\phi\alpha/2)\over \tanh(M\phi/2)},~~
E=E_0+{J\sinh \phi\sinh M\phi \over \cosh M\phi-\cos\phi\alpha}+2Mh.
\end{equation}
Thus the dispersion is
\begin{equation}
E=E_0+2Mh+J\sinh\phi\Bigl[{\cosh M\phi-\cos K\over \sinh M\phi}\Bigr].
\label{eq:xxzeng}
\end{equation}
This excitation energy gives (\ref{eq:xxxeng}) in the limit $\Delta\to 1$.
In the limit of large $\Delta$ the energy is $J\Delta+2Mh$. This is $M$
successive down spins in a sea of up spins.
The lowest energy state at $\Delta>1$ with $M$ down spins is the $M$ string
state given by (\ref{eq:xxzeng}) with zero total momentum. So 
the energy of the ground state for $\Delta>1$ is
\begin{equation}
E=-{JN\Delta\over 4}-(N-2M)h+J\sinh\phi\tanh{M\phi\over 2}.
\end{equation}

\section{Thermodynamic equations for the XXZ model for $\Delta>{\it 1}$}
Gaudin derived a set of thermodynamic Bethe ansatz equations 
at $\Delta>1$ (\cite{gaudin71}). 
The wave function for $M$ down spins in the infinite lattice is
\beqa
&&f(n_1,n_2,...,n_M)=(z_1z_2..z_M)^{n_1}\sum_PA(P)
\prod_{j=2}^{M}(\prod^M_{l=j}z_{Pl})^{n_{j+1}-n_j},\nonumber\\
&&z_j=e^{ik_j}=({\sin{\phi\over 2}(x_j+i)\over \sin{\phi\over 2}(x_j-i)}).
\label{eq:wfxxz}\enqa
\index{normalizability condition}
This corresponds to (\ref{eq:wfxxx}). 
From the normalizability condition of the wave function we have
\beqa
A(I)\ne 0, A(P\ne I)=0 ,\quad\vert z_1z_2...z_M\vert=1\nonum\\
|\prod_{l=j+1}^Mz_l|\le1,~~j=1,2,3,...,M-1.
\enqa
These conditions are satisfied only if
\beq
x_j=\alpha+(M+1-2j)i,\quad j=1,2,...,M, \quad Q\ge\alpha>-Q.
\enq
We can show that
\beqa
|\prod_{l=j}^Mz_l|=|{\sin{\phi\over 2}(\alpha+i(M-2j))\over
\sin{\phi\over 2}(\alpha-i M)}|=\sqrt{\cosh \phi(M-2j)-\cos \phi\alpha
\over \cosh \phi n-\cos \phi\alpha}\le 1, \nonum\\
~~1\le j\le M-1,\nonum
\enqa
for arbitrary $M$.  
Thus a string with arbitrary length is possible for the 
XXZ chain at $|\Delta|>1$. 
In the case $|\Delta|<1$ the string condition is more complicated than the 
case $|\Delta|\ge 1$.
From (\ref{eq:xxzbethe}) we have the following equation corresponding to 
(\ref{eq:xxxstring}),
\beq
e_n^N(x^n_\alpha)=\prod_{j=1}^n e^N(x_\alpha^{n,j})=\prod_{(m,\beta)\ne
(n,\alpha)}E_{nm}(x_\alpha^n-x_\beta^m).
\label{eq:xxzstring0}\enq
Here 
\beq
e_n(x)={\sin{\phi\over 2}(x+i n )\over \sin{\phi\over 2}(x-i n)},
\enq
\beq
E_{nm}(x)\equiv\left\{\matrix{
e_{\vert n-m\vert}(x)e_{\vert n-m\vert+2}^2(x)
e^2_{\vert n-m\vert+4}(x)...e^2_{n+m-2}(x)e_{n+m}(x)\cr
{\rm for}\quad n\ne m,\cr
e^2_{2}(x)e^2_{4}(x)...e^2_{2n-2}(x)e_{2n}(x)\quad{\rm for}
\quad n=m.}\right.
\enq
$x^n_\alpha$ is the real part of $\alpha$-th string in the strings 
of length $n$. 
The logarithm of (\ref{eq:xxzstring0}) is
\beq
N\theta_n(x_\alpha^n)=2\pi I_\alpha^n+\sum_{(m,\beta)\ne
(n,\alpha)}\Theta_{nm}(x_\alpha^n-x_\beta^m),\label{eq:xxzstring}
\enq
where
$$\theta_n(x)=2\tan^{-1}({\tan{x\phi\over 2}\over \tanh{n\phi\over 2}})
+2\pi[{x+Q\over 2Q}],$$
and 
\beq
\Theta_{nm}(x)\equiv\left\{\matrix{\theta_{\vert n-m\vert}(x)
+2\theta_{\vert n-m \vert +2}(x)+...+2\theta_{n+m-2}(x)
+\theta_{n+m}(x)\cr
{\rm for}\quad n\ne m,\cr
2\theta_{2}(x)+2\theta_{4}(x)+...+2\theta_{2n-2}(x)
+\theta_{2n}(x)\quad {\rm for}\quad n=m.}\right.
\enq
The function $\theta_n(x)$ is a quasi periodic function which satisfies
$$\theta_n(x+2jQ)=\theta_n(x)+2\pi j,~~j={\rm integer}.$$
We consider the energy of general eigenstates which is 
given by the set of quantum numbers $\{I_\alpha^n\}$,
\beq
E(\{I_\alpha^n\})=N(-h-{J\Delta\over 4})+\sum_{n,\alpha}
({2\pi J\sinh \phi\over \phi} {\bf a}_n(x_\alpha^n)+2hn),
\enq
where 
\beq
{\bf a}_n(x)={1\over 2\pi}{\phi\sinh n\phi\over \cosh n\phi-\cos\phi x}.
\label{eq:funcan} 
\enq
The partition function of the XXZ model is as follows,
\beq
{\cal Z}=\sum_{M=0}^{[N/2]}(1+(1-\delta_{N,2M})\exp-{(N-2M)h\over T})
\sum_{\{I_\alpha^n\}}\exp[-T^{-1}E(\{I_\alpha^n\})].
\enq
Corresponding to (\ref{eq:xt35}) we define the following functions,
\beq
h_n(x)\equiv \theta_n(x)-N^{-1}\sum_{(m,\alpha)}\Theta_{nm}(x-x^m_\alpha).
\enq
Using this function we can determine the position of holes for n-strings. 
We define the distribution functions of particles and holes of n-strings as 
$\rho_n(x)$ and $\rho_n^h(x)$. By the equation (\ref{eq:xxzstring}) we have 
the conditions for these two kinds of functions
\beq
{\bf a}_n(x)=\rho_n(x)+\rho^h_n(x)+\sum_m{\bf T}_{nm}*\rho_m(x).
\label{eq:xxzdis}
\enq 
Here ${\bf a}_n$ was defined in (\ref{eq:funcan}) and 
\beq
{\bf T}_{nm}(x)\equiv
\left\{\matrix{{\bf a}_{|n-m|}(x)+2{\bf a}_{|n-m|+2}(x)+2{\bf a}_{|n-m|+4}(x)
+...
\cr+2{\bf a}_{n+m-2}(x)+{\bf a}_{n+m}(x)~~~{\rm for}\quad n\ne m,\cr 
2{\bf a}_{2}(x)+2{\bf a}_{4}(x)+...
+2{\bf a}_{2n-2}(x)+{\bf a}_{2n}(x)\quad {\rm for}\quad n=m.}\right.\enq   
Here the meaning of convolution of two periodic functions ${\bf a}$ and 
${\bf b}$ with periodicity $2Q$ is redefined 
\beq
{\bf a}*{\bf b}(x)\equiv\int^Q_{-Q}{\bf a}(x-y){\bf b}(y)dy,
\enq
The energy per site is
\beq
e=-({J\Delta\over 4}+h)
+\sum_{n=1}^\infty\int^Q_{-Q}{\bf g}_n(x)\rho_n(x)dx,$$
$${\bf g}_n(x)\equiv {2\pi J\sinh \phi\over \phi}{\bf a}_n(x)+2nh.
\enq
The entropy per site ${\it s}$ is
\beq
{\it s}=\sum^\infty_{n=1}\int^Q_{-Q} 
\rho_n(x)\ln(1+{\rho_n^h(x)\over\rho_n(x)})
+\rho_n^h(x)\ln(1+{\rho_n(x)\over\rho_n^h(x)})dx.
\enq 
The condition of minimizing the free energy $e-T{\it s}$ yields equations 
for $\eta_n(x)\equiv\rho_n^h(x)/\rho_n(x)$,
\beq
\ln\eta_n(x)={{\bf g}_n(x)\over T}
+\sum^\infty_{m=1}{\bf T}_{nm}*\ln(1+\eta_m^{-1}(x)).
\enq
This set of equations is equivalent to the following one
\beqa
&&\ln\eta_1(x)={2\pi J\sinh\phi \over T\phi}{\bf s}(x)+{\bf s}
*\ln(1+\eta_2(x)),\label{eq:gteq4a}\\
&&\ln\eta_n(x)={\bf s}*\ln(1+\eta_{n-1}(x))(1+\eta_{n+1}(x)),
\label{eq:gteq4b}\\
&&\lim_{n\to \infty}{\ln\eta_n\over n}={2h\over T},\label{eq:gteq4c}
\enqa
where
\beq
{\bf s}(x)={1\over 4}\sum_{n=-\infty}^\infty 
{\rm sech}({\pi (x-2nQ)\over 2}).
\enq
The free energy per site is 
\beq
f=-({J\Delta\over 4}+h)-T\sum_{n=1}^\infty\int^Q_{-Q}
{\bf a}_n(x)\ln(1+\eta_n^{-1}(x))dx.
\enq
Corresponding to (\ref{eq:xtfe2}) we have another 
expression for the free energy,
\beq
f=J\Bigl[{2\pi\sinh \phi\over \phi}\int^Q_{-Q} {\bf a}_1(x){\bf s}(x)dx
-{\Delta\over 4}\Bigr]
-T\int^Q_{-Q}{\bf s}(x)\ln(1+\eta_1(x))dx.
\enq
(\ref{eq:xt59a}-\ref{eq:xt59c}) and (\ref{eq:gteq4a}-\ref{eq:gteq4c}) have 
the almost same structure. These equations are called Gaudin-Takahashi 
equation (\cite{gaudin71}, \cite{tak71a}).
\index{Gaudin-Takahashi equation}

\section{Theory for $|\Delta|<{\it 1}$ XXZ model}
\subsection{String solution of an infinite XXZ model with $|\Delta|<{\it 1}$}
The shapes of strings for $\vert\Delta\vert<1$ are quite different from those
at $\vert\Delta\vert\ge 1$.
Takahashi and Suzuki proposed a condition of the strings and
constructed thermodynamic integral equations (\cite{TS72}).  
Later Hida,
Fowler and Zotos derived these conditions from the normalizability
condition of the string wave function for an 
infinite chain (\cite{hida81}, \cite{f&z81}).
\index{normalizability condition}
For $ \vert \Delta\vert< 1$ there are two kinds of strings, one of 
which has the center on the real axis and the other is centered on the 
$p_0 i$ axis,
\beqa
x_j&=&\alpha+(n+1-2j)i,~~j=1,2,...,n,\label{eq:string1}\\
x_j&=&\alpha+(n+1-2j)i+p_0 i,~~j=1,2,...,n.\label{eq:string2}
\enqa
We designate that the string of the former type has parity $v=1$ and 
that the latter has parity $v=-1$. 
Applying the normalizability condition of the form (\ref{eq:string1}) yields 
$$1>|\prod_{l=j+1}^n z_l|=|{\sinh{\gamma\over 2}(\alpha+i(n-2j))\over 
\sinh{\gamma\over 2}(\alpha-in)}|=\sqrt{\cosh \gamma\alpha-\cos \gamma(n-2j)
\over \cosh \gamma\alpha-\cos \gamma n}.$$
Thus $\cos \gamma n<\cos \gamma(n-2j)$ for $j=1,2,3,...,n-1$. 
For (\ref{eq:string2}) we have
$$1>|\prod_{l=j+1}^n z_l|=|{\cosh{\gamma\over 2}(\alpha+i(n-2j))\over 
\cosh{\gamma\over 2}(\alpha-in)}|=\sqrt{\cosh \gamma\alpha+\cos \gamma(n-2j)
\over \cosh \gamma\alpha+\cos \gamma n},$$
and therefore $\cos \gamma n>\cos \gamma(n-2j)$ for $j=1,2,3,...,n-1$.
Then from the normalizability condition we get
\beqa
&&0<v(\cos((n-2j)\gamma)-\cos(n\gamma))\nonum\\
&&=2v\sin((n-j)\gamma)\sin(j\gamma),~~{\rm for}~~
j=1,2,...,n-1.\label{eq:norm}
\enqa
This is equivalent to
\beqa
(-1)^{[{(n-j)\gamma\over\pi}]+ [{j\gamma\over\pi}]}=v,~~
{\rm for} ~~j=1,2,...,n-1,\label{eq:norm2}\\
{j\gamma\over \pi}\ne [{j\gamma\over \pi}],~~
{\rm for} ~~j=1,2,...,n-1,\label{eq:norm3}
\enqa
where $[x]$ denotes the maximum integer less than or equal to $x$ (Gauss' 
symbol). 
For rational $p_0=\pi/\gamma$, (\ref{eq:norm3}) is a strong condition. 
If $p_0=n_1/n_2$, and $n_1$ and $n_2$ 
are coprime, the string with length greater than $n_1$ cannot satisfy 
at least one of (\ref{eq:norm3}). Thus $n\ge n_1+1$ strings are forbidden. 
Moreover for a $n=n_1$ string, the momentum is always $0$ or $\pi$. 
So this string 
has also no meaning for the thermodynamics. Next we seek 
the number $n$ and parity $v$ which satisfies (\ref{eq:norm2}) within 
$n<n_1$. Equation (\ref{eq:norm2}) is equivalent to        
$$[{(n-j)\gamma\over\pi}]+ [{j\gamma\over\pi}]\equiv 
[{(n-j-1)\gamma\over\pi}]+ [{(j+1)\gamma\over\pi}] ({\rm Mod}2),
~~j=1,2,...,n-2,$$
$$ [{(n-1)\gamma\over\pi}]\equiv {1-v\over 2}~~({\rm Mod}2).$$
As $[{(n-j)\gamma\over\pi}]-[{(n-j-1)\gamma\over\pi}]$ is $0$ or $1$ and 
$[{j\gamma\over\pi}]-[{(j+1)\gamma\over\pi}]$ is $0$ or $-1$, we obtain
$$[{(n-j)\gamma\over\pi}]+ [{j\gamma\over\pi}]= 
[{(n-j-1)\gamma\over\pi}]+ [{(j+1)\gamma\over\pi}],
~~j=1,2,...,n-2.$$
These are strong restrictions on the parity $v$ 
and the length of the string $n$. \index{parity of string}
The above conditions are equivalent to
the following conditions which were given in (\cite{TS72}).  
The length $n$ of a string  should satisfy
\beqa
2\sum^{n-1}_{j=1}[j\gamma/\pi]=(n-1)[(n-1)\gamma/\pi],\label{eq:cond1}\\
v\sin\{(n-1)\gamma\}\ge 0.\label{eq:cond2}
\enqa 
This condition was first introduced 
under the assumption 
that these strings form a complete half-filled state (\cite{TS72}).  
Later Hida, Fowler and Zotos showed that conditions 
(\ref{eq:cond1},\ref{eq:cond2}) can be rederived from the 
normalizability conditions of the wave function for $N\to \infty$ and 
finite $M$ (\cite{hida81}, \cite{f&z81}).  
For a given value of $\Delta$ (or $\gamma$) we can determine the
series of $n$ which satisfies the conditions 
(\ref{eq:cond1}) and (\ref{eq:cond2}). If $\gamma/\pi$ is a rational number, 
this series becomes finite and the number of unknown functions is also finite.  
We consider the $\gamma=\pi/\nu,~\nu=$ integer case.
In this case $n=1,2,...,\nu$ satisfy (\ref{eq:cond1}). 
For $n=1$ both 
$v=1$ and $-1$ are possible. For $n=2,3,...,\nu-1$, only $v=+1$ states are 
possible. These excitations have the following energy and momentum
\beqa
E&=&-2J {\sin \gamma \sin (n \gamma)\over v\cosh(\alpha \gamma)
-\cos n\gamma}+2nh,\\
K&=&-i\ln\Bigl(-
{\sinh\frac{1}{2}(\alpha \gamma+i(1-v)\pi/2+in\gamma)
\over \sinh\frac{1}{2}(\alpha \gamma+i(1-v)\pi/2-in\gamma)}\Bigr).
\enqa
The energy and momentum have the following relation,
$$E=-J\sin\gamma{\cos n\gamma-\cos K\over \sin n\gamma}+2nh.$$
The momentum is restricted to the region
$$|K|<\pi-(n\gamma-\pi[{n\gamma\over\pi}])~~{\rm for}~~v=1,$$
$$\pi\ge |K|>\pi-(n\gamma-\pi[{n\gamma\over\pi}])~~{\rm for}~~v=-1.$$
Then for $n=\nu$, the energy and momentum are always zero. Only 
one state is obtained from this string solution. 
So we exclude this $n=\nu$ state from the thermodynamics of this case. 
So $\nu$ string states $(1,+),(2,+),...,(\nu-1,+),(1,-)$ play important 
roles. Especially at $\Delta=0,~ \gamma=\pi/2,~ \nu=2$ we have only 
string states $(1,+),(1,-)$. These are single states at momentum $|K|<\pi/2$ 
and $|K|>\pi/2$.  

Next we consider the $\gamma=\pi/(\nu_1+1/\nu_2)$ case.
$(1,+),(2,+),...,(\nu_1-1,+),$
$(1,-),(1+\nu_1,+),(1+2\nu_1,-),...
(1+(\nu_2-1)\nu_1,(-1)^{\nu_2-1}),(\nu_1,(-1)^{\nu_2})$ 
satisfy conditions (\ref{eq:cond1}, \ref{eq:cond2}). 
Thus $\nu_1+\nu_2$ strings are necessary to describe the 
thermodynamics of this case. 

For a general rational number between 0 and 1, we can express it 
by a continued fraction with length $l$, \index{continued fraction}
\beq
{\gamma\over\pi}={1|\over |\nu_1}+{1|\over |\nu_2}+...+{1|\over |\nu_l},
~~\nu_1,\nu_2,...,\nu_{l-1}\ge1,~~\nu_l\ge 2.
\enq 
We define the following series of numbers $y_{-1},y_0,y_1,...,y_l$ and 
\par\noindent
$m_0,m_1,...,m_l$ as
\beqa &&y_{-1}=0,~~ y_0=1,~~ y_1=\nu_1~~{\rm  and}
~~ y_i=y_{i-2}+\nu_iy_{i-1},\nonum\\
&&m_0=0,~~ m_i=\sum^i_{k=1}\nu_k.
\enqa
The general rule to determine the parity $v$ and length $n$ is as follows
\beqa
&&n_j=y_{i-1}+(j-m_i)y_i,~~v_j=(-1)^{[(n_j-1)/p_0]}
~~{\rm for}~~m_i< j<m_{i+1},\nonum\\
&&n_{m_l}=y_{l-1},~~v_{m_l}=(-1)^l. 
\enqa
The number of strings is $m_l$. We give examples for some rational numbers 
in Tables (\ref{tab:t61}), (\ref{tab:t62}). We put $x_j^\alpha$ as the 
real part of strings with parity $v_j$ and length $n_j$. $\alpha$ takes 
values from 1 to $M_j$. 
We find the following relations for these series of numbers
\beqa
&&n_j={1\over 2}[(1-2\delta_{m_i,j})n_{j-1}+n_{j+1}],~~{\rm for}~~
m_i\le j\le m_{i+1}-2,\nonum\\
&&n_j=(1-2\delta_{m_{i-1},j})n_{j-1}+n_{j+1},~~{\rm for}~~
 j= m_i-1,~~i<l,\nonum\\
&&n_0=0,~~n_{m_l}+n_{m_l-1}=y_l.\label{eq:nj1}
\enqa

\begin{table}
\caption{Length $n_j$, parity $v_j$ and $q_j$ of strings for 
some rational values of $\gamma/\pi$}

\begin{tabular*}{120mm}{@{\extracolsep{\fill}}|r||rrr|rrr|rrr|} \hline
$j$
&${1\over 5}$&$ $&$ $  & ${3\over 16}$ &$=$ & ${1\over 5+{1\over 3}} $ & 
${13\over 69}$ &$=$ &${1\over 5+{1\over 3+{1\over 4}}}$ \\ \hline
 1  & $1$&$~+$ &$~~4$  & $1$ &$~+$ & $~13/3$   & $ 1$ &$~+$& $~~56/13 $                            \\
 2  & $2$ &$~+$&$~~3$  & $2$&$~+$&$~10/3$   & $ 2$&$~+$&$~~43/13 $                            \\
 3  & $3$&$~+$&$~~2$  & $3$&$~+$&$~~7/3 $   & $ 3$&$~+$&$~~30/13 $                            \\
 4  & $4$&$~+$&$~~1$  & $4$&$~+$&$~~4/3 $   & $ 4$&$~+$&$~~17/13 $                            \\
 5  & $1$&$~-$&$~-1$ & $1$&$~-$&$~-3/3$   & $ 1$&$~-$&$~-13/13 $                            \\
 6  &  & &       & $6$&$~+$&$~-2/3$   & $ 6$&$~+$&$~~-9/13 $                            \\
 7  &  & &        & $11$&$~-$&$~-1/3$  & $11$&$~-$&$~~-5/13 $                            \\
 8  &  & &        & $5$&$~+$&$~~1/3 $   & $ 5$&$~+$&$~~~4/13 $                            \\
 9  &   & &       &   &  &         & $21$&$~-$&$~~~3/13 $                            \\
 10  &  & &      &    & &          & $37$&$~+$&$~~~2/13 $                            \\
 11  &  & &       &   & &           & $53$&$~-$&$~~~1/13 $                            \\
 12  &  & &      &   & &           & $16$&$~-$&$~~-1/13 $                            \\
\hline
\end{tabular*}
\label{tab:t61}
\end{table}

\begin{table}
\caption{Length $n_j$, parity $v_j$ and $q_j$ of strings 
for conjugate values of $\gamma/\pi$ in previous table}

\begin{tabular*}{120mm}{@{\extracolsep{\fill}}|r||rrr|rrr|rrr|} \hline
$j$ 
&${4\over 5}$&$=$&${1\over 1+{1\over 4}}$  & ${13\over 16}$&$
=$&${1\over 1+{1\over {4+{1\over 3}}}} $ & 
${56\over 69}$&$=$&${1\over {1+{1\over 4+{1\over 3+{1\over 4}}}}}$ \\ \hline
 1  & $1$&$~-$&$~-4/4$     & $1$&$~-$&$~-13/13$       &  $1$&$~-$&$~-56/56 $                \\
 2  & $2$&$~+$&$~-3/4$     & $2$&$~+$&$~-10/13$       &  $2$&$~+$&$~-43/56 $                \\
 3  & $3$&$~-$&$~-2/4$     & $3$&$~-$&$~~-7/13$       &  $3$&$~-$&$~-30/56 $                \\
 4  & $4$&$~+$&$~-1/4$     & $4$&$~+$&$~~-4/13$       &  $4$&$~+$&$~-17/56 $                \\
 5  & $1$&$~+$&$~~1/4$     & $1$&$~+$&$~~~3/13$       &  $1$&$~+$&$~~13/56 $                \\
 6  &   &    &         & $6$&$~+$&$~~~2/13$       &  $6$&$~+$&$~~9/56 $                \\
 7  &  & &            &$11$&$~+$&$~~~1/13$       & $11$&$~+$&$~~~5/56 $                 \\
 8  & &   &            & $5$&$-$&$-1/13$       &  $5$&$~-$&$~~-4/56 $                \\
 9  & &  &            &   & &                 & $21$&$~+$&$~~-3/56 $            \\
 10  & & &              &  & &                  & $37$&$~-$&$~~-2/56 $            \\
 11  &   & &            & & &                   & $53$&$~+$&$~~-1/56 $            \\
 12  &  & &             &   & &                 & $16$&$~+$&$~~~1/56 $            \\
\hline
\end{tabular*}
\label{tab:t62}
\end{table}

\newpage

\begin{figure}
\par\noindent
a) $\gamma=\pi/5$\hskip 4.5truecm b) $\gamma=4\pi/5$ \par\noindent      
\epsfxsize=6cm \epsfbox{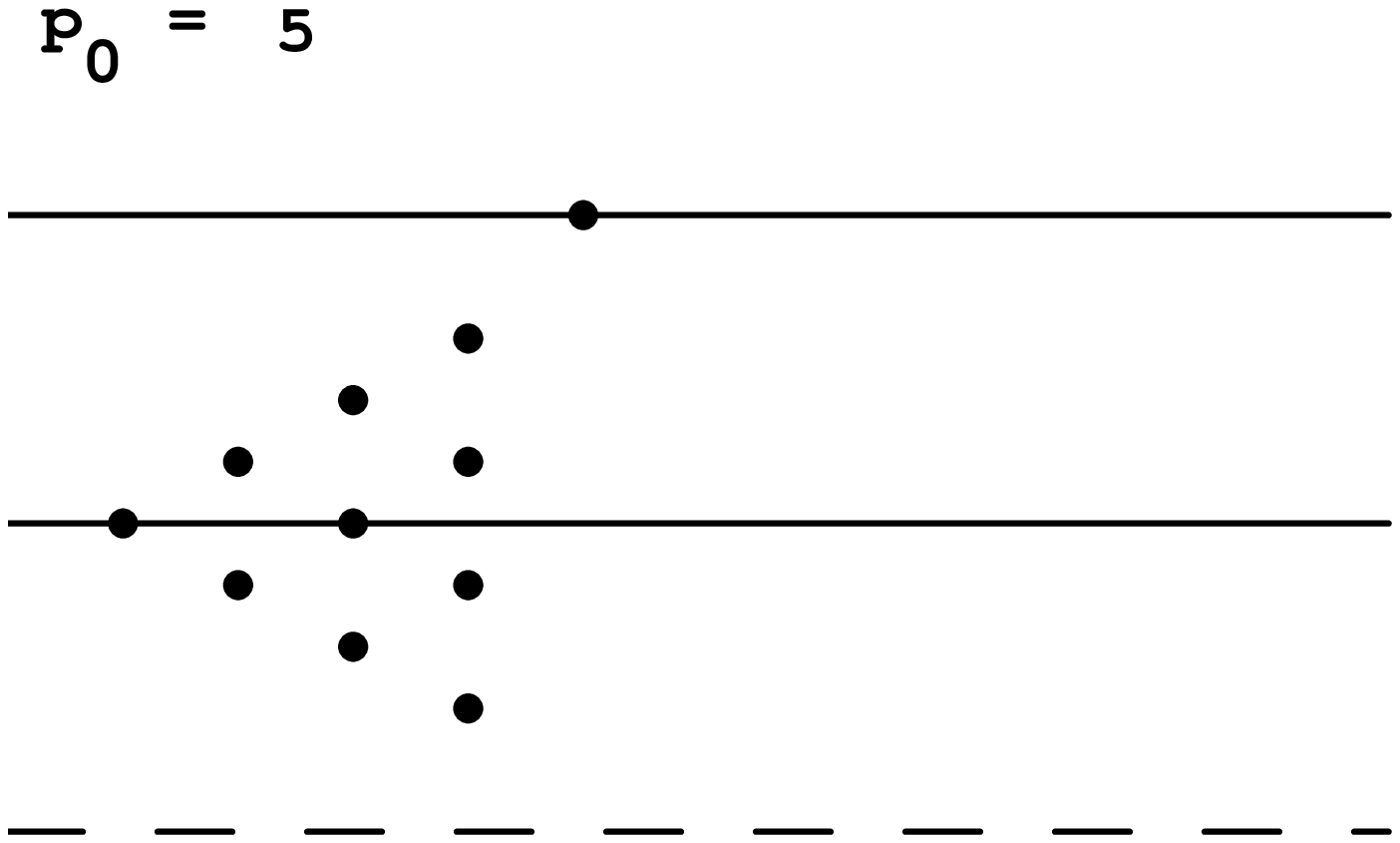}~~\epsfxsize=6cm \epsfbox{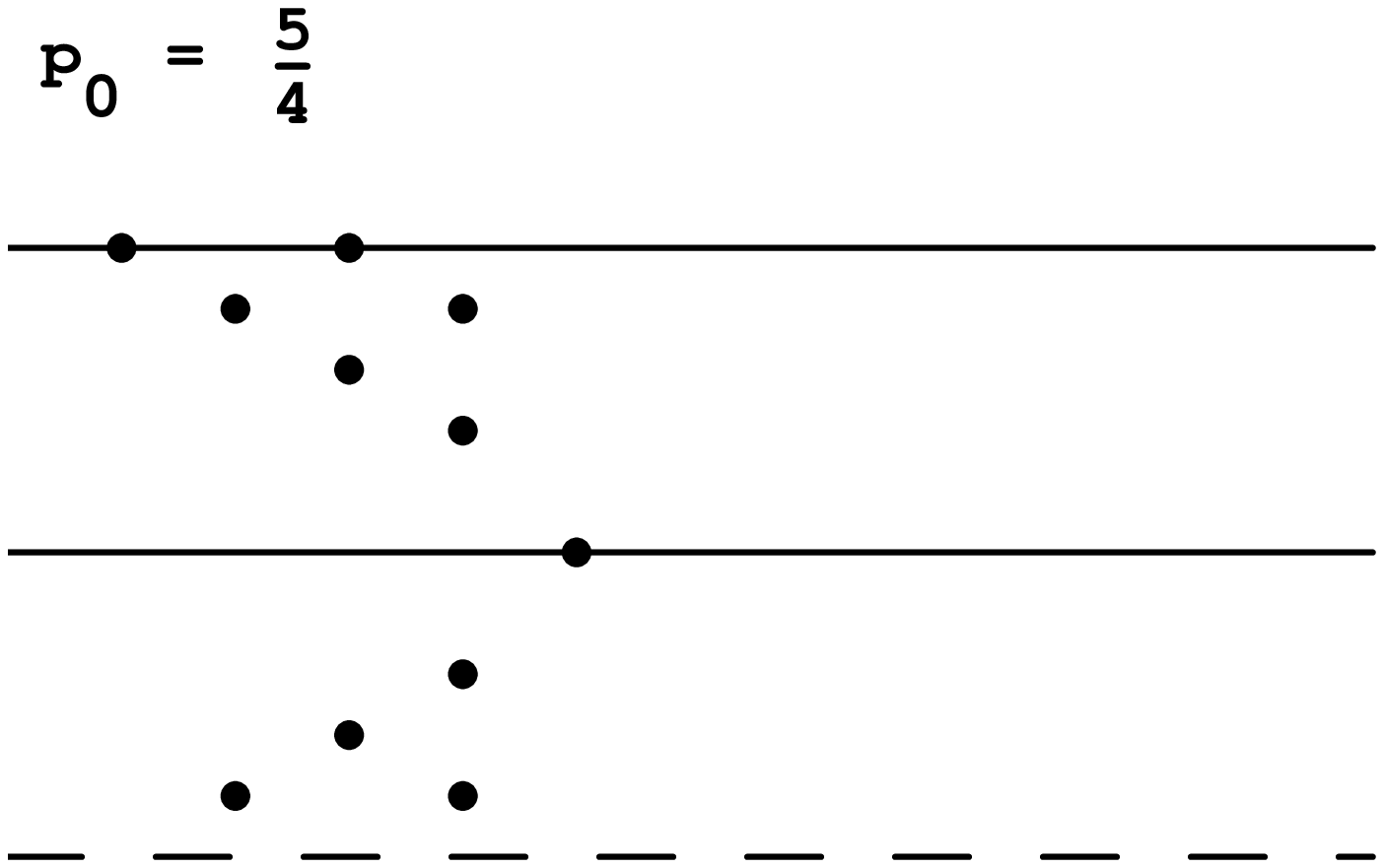}
\par\noindent
\vskip .8truecm
c) $\gamma=3\pi/16$\hskip 4.3truecm d) $\gamma=13\pi/16$ \par\noindent      
\epsfxsize=6cm \epsfbox{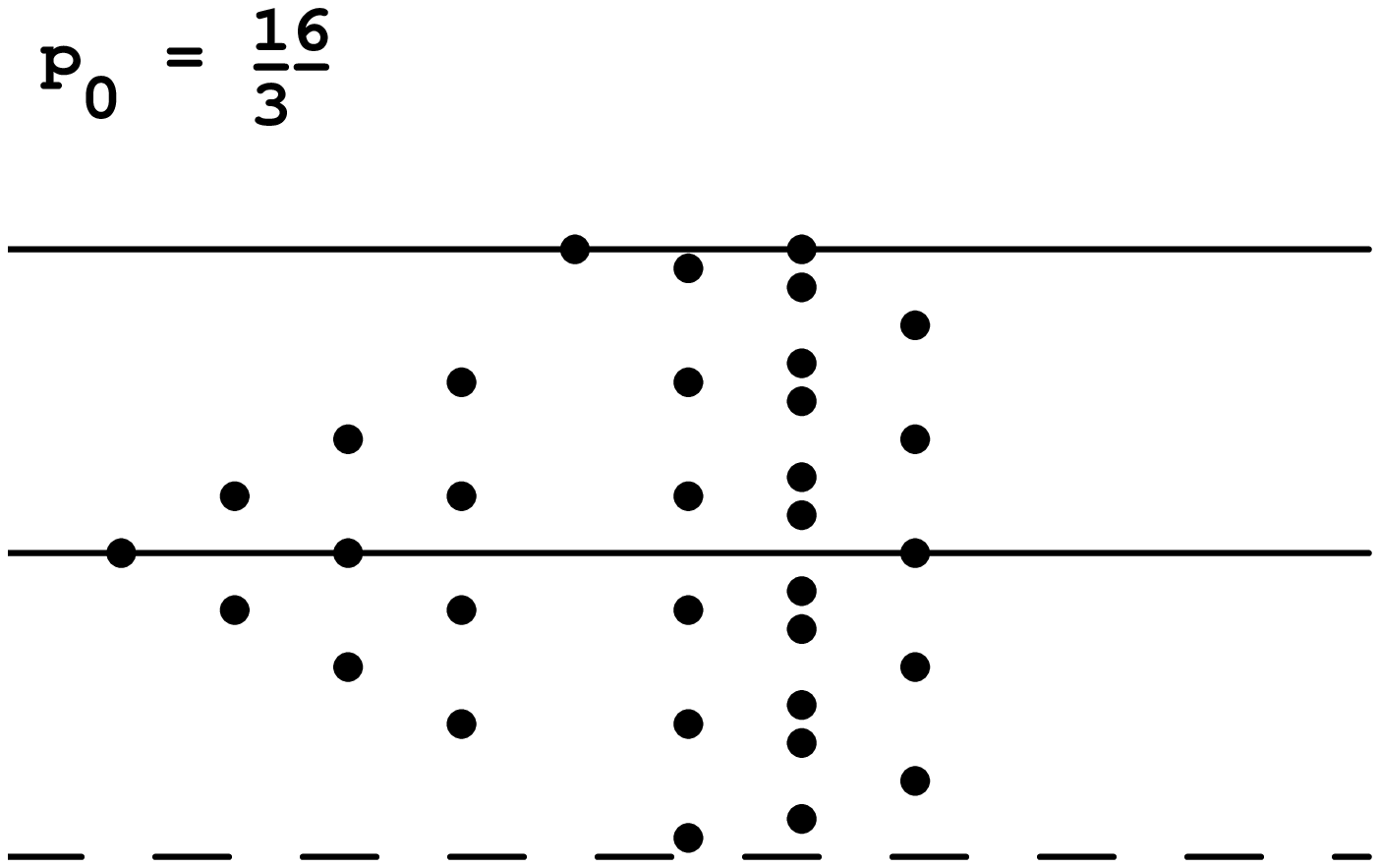}~~\epsfxsize=6cm \epsfbox{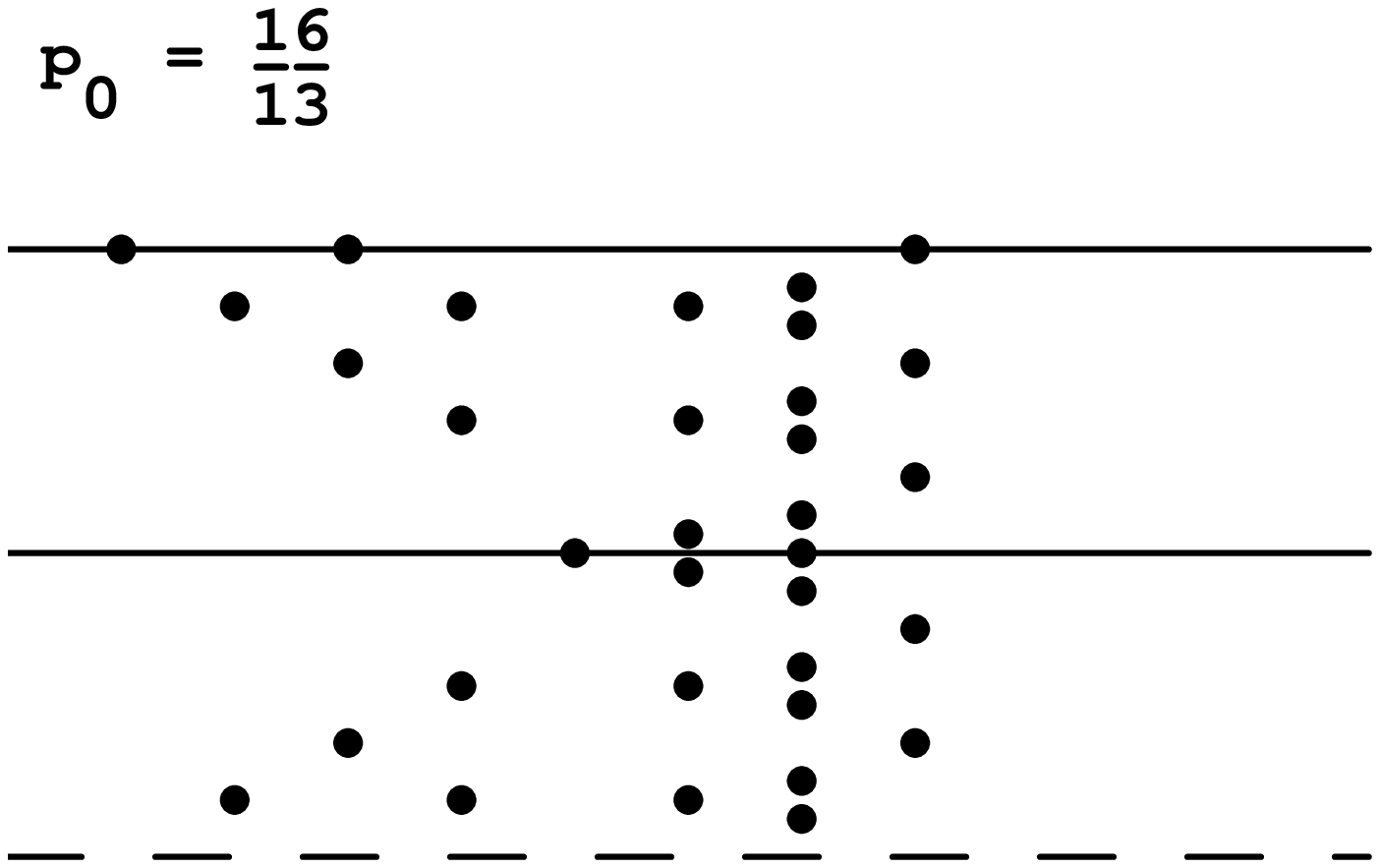}
\par\noindent
\vskip .8truecm
e) $\gamma=13\pi/69$\hskip 4.0truecm f) $\gamma=56\pi/69$ \par\noindent      
\epsfxsize=6cm \epsfbox{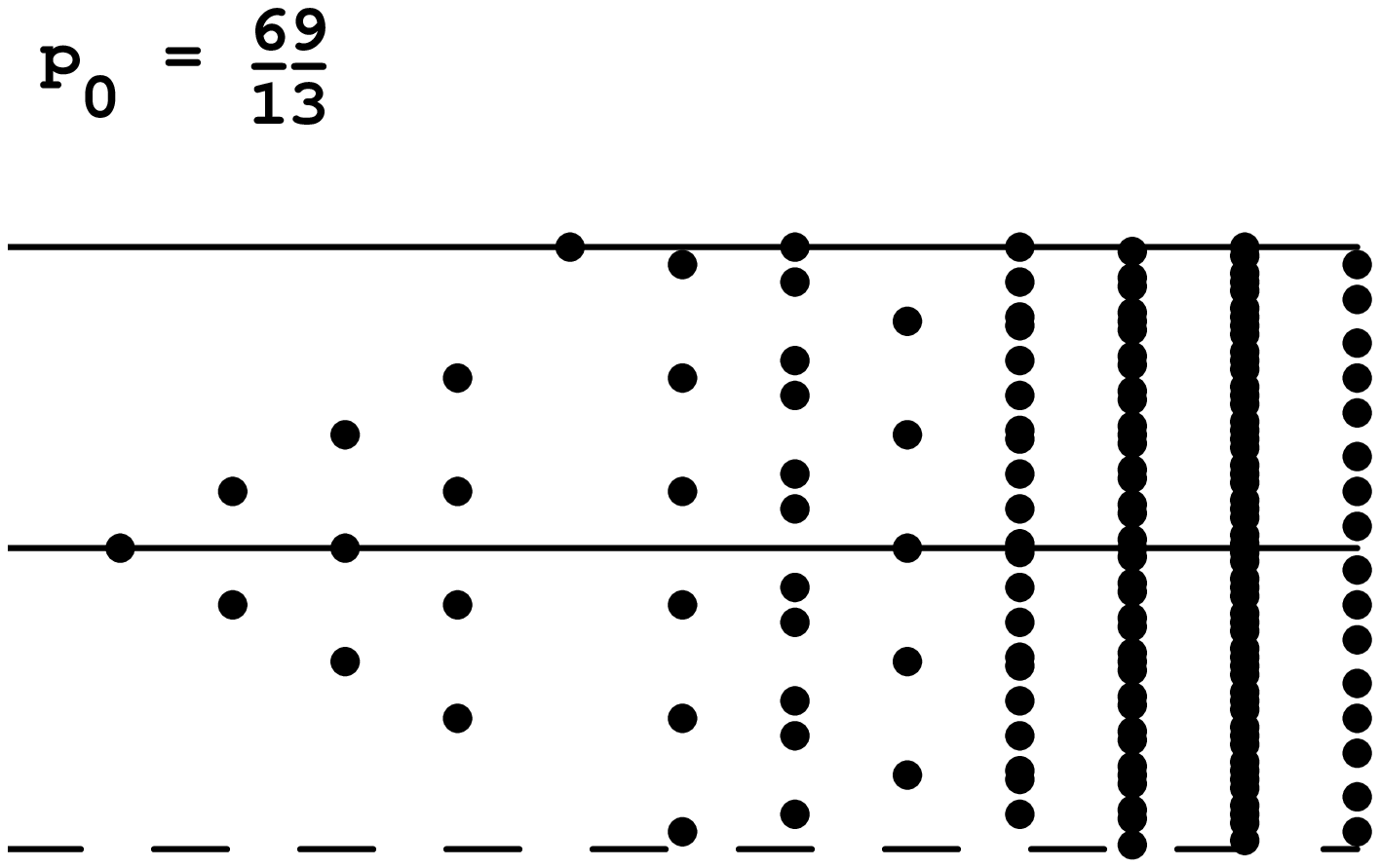}~~\epsfxsize=6cm \epsfbox{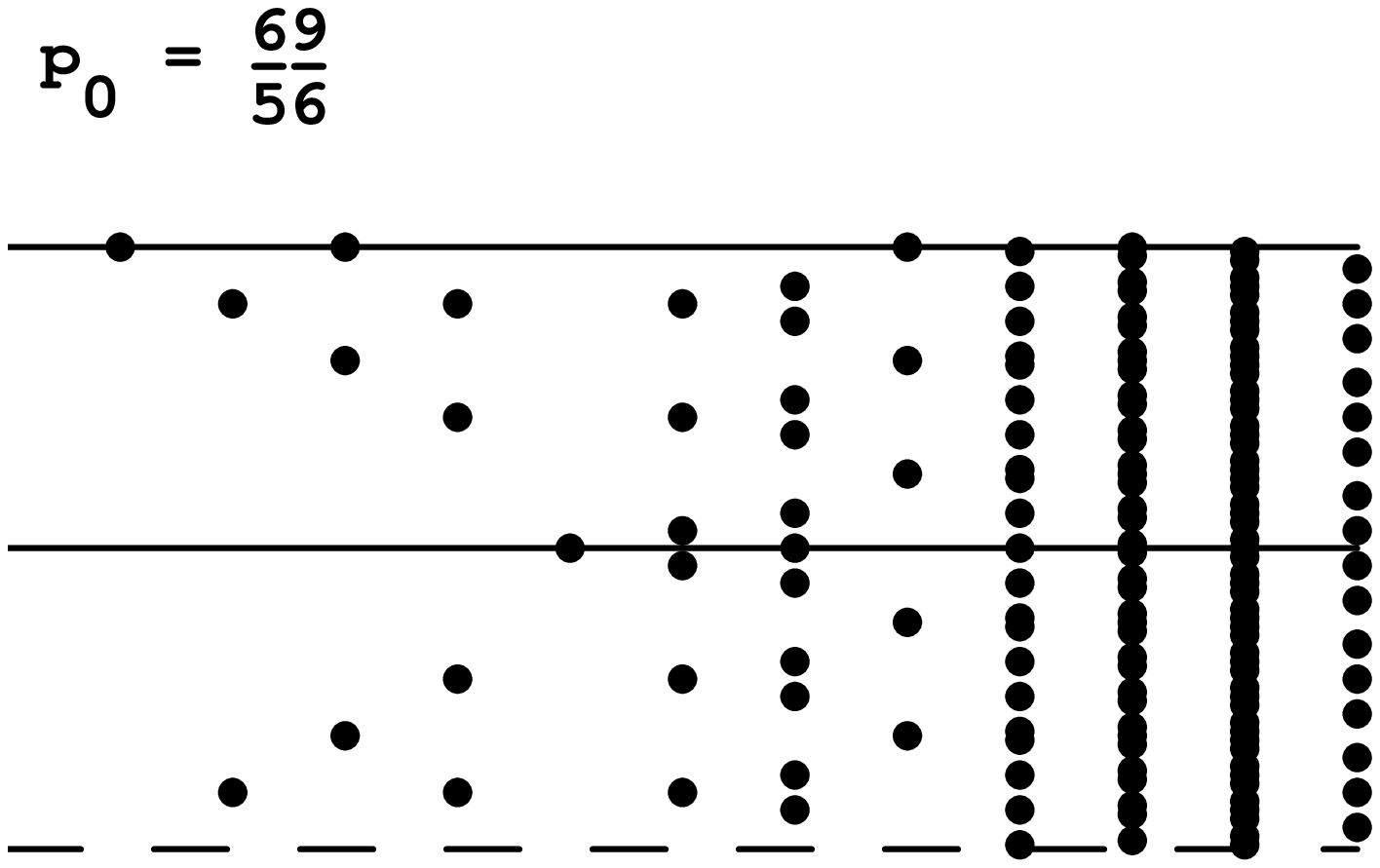}
\vskip .8truecm
\caption{
Strings on the complex plane for given values of $p_0=\pi/\gamma$.
If we change $\gamma\to \pi-\gamma$, the strings are almost the same. But 
imaginary parts of $\gamma x_j$ shift by $\pi$.  
}
\end{figure}

\subsection{Scattering phase shift among strings}
Corresponding to (\ref{eq:xxxstring}) 
and (\ref{eq:xxzstring0}) we
have Bethe ansatz equations for strings as follows
\beq
\{e_j(x_\alpha^j)\}^N=-\prod_{k=1}^{m_l}\prod_{\beta=1}^{M_k}
E_{j,k}(x_\alpha^j-x_\beta^k),\label{eq:xylike1}
\enq
where
\beq
e_j(x)=g(x;n_j,v_j),
\enq

\beq
E_{j,k}(x)=\left\{\matrix{
g(x;2n_j,v_jv_k)\prod_{l=1}^{n_j-1}g^2(x;2l,v_jv_k)~~{\rm for}~~n_j=n_k,\cr
g(x;(n_j+n_k),v_jv_k)g(x;|n_j-n_k|,v_jv_k)\cr
\times \prod_{l=1}^{Min(n_j,n_k)-1}g^2(x;|n_j-n_k|+2l,v_jv_k)
~~{\rm for}~~n_j\ne n_k,}\right.
\enq
\beq
g(x;n,+)={\sinh{ \gamma\over 2}(x+in)\over \sinh{ \gamma\over 2}(x-in)}, ~~
g(x;n,-)=-{\cosh{ \gamma\over 2}(x+i n)\over \cosh{ \gamma\over 2}(x-in)}.
\enq
The logarithm of (\ref{eq:xylike1}) yields
\beq
N\theta_j(x_\alpha^j)=2\pi I_\alpha^j+\sum_{k=1}^{m_l}\sum_{\beta=1}^{M_k}
\Theta_{j,k}(x_\alpha^j-x_\beta^k),~~\alpha=1,2,...,M_j,
\enq
where 
\beqa
\theta_j(x)=f(x;n_j,v_j),~~\Theta_{jk}(x)=f(x;|n_j-n_k|,v_jv_k)+\nonum\\
f(x;n_j+n_k,v_jv_k)
+2\sum_{i=1}^{Min(n_j,n_k)-1}
f(x;|n_j-n_k|+2i,v_jv_k),
\enqa
and
\beq
f(x;n,v)\equiv\left\{\matrix{0~~~~~~~~{\rm for}~~ n/p_0={\rm integer},\cr
2v\tan^{-1}\{(\cot(n\pi/2p_0))^v\tanh( \pi x/2p_0)\}~~{\rm otherwise}.}\right.
\label{eq:funcf}
\enq
The quantity $I_\alpha^j$ is an integer (half-odd integer) for 
$M_j$ odd (even), which is located in the region
\beq
|I_\alpha^j|<{1\over 2\pi}|N\theta_j(\infty)-\sum_{k=1}^{m_l}
M_k\Theta_{j,k}(\infty)|.
\enq
The function $f(x;n_j,v_j)$ is a monotonically increasing function for 
$m_{2i}\le j<m_{2i+1}$ and a monotonically decreasing function for 
$m_{2i-1}\le j<m_{2i}$. 

\subsection{Bethe-ansatz equation for XXZ model with $|\Delta|<{\it 1}$}
Following Yang and Yang we define particles and holes of strings. We obtain 
an integral equation for distribution functions $\rho_j$ and $\rho_j^h$ 
of particles and holes of strings in the thermodynamic limit,
\beqa
&&a_j(x)=\sign(q_j)(\rho_j(x)+\rho_j^h(x))
+\sum_{k=1}^{m_l}T_{j,k}*\rho_k(x).
\label{eq:xylinear}
\enqa
Here
\beq
a_j(x)\equiv (2\pi)^{-1}{d\over dx}\theta_j(x),~~ T_{j,k}(x)\equiv(2\pi)^{-1}
{d\over dx}\Theta_{j,k}(x).\label{eq:xysym}
\enq
The symbol $a*b$ denotes the convolution of $a(x)$ and $b(x)$ as follows
\beq
a*b(x)=\int^\infty_{-\infty}a(x-y)b(y)dy.
\enq
\index{convolution}
The functions $a_j(x)$ and their Fourier transforms 
$\tilde{a}_j(\omega)$ are written as
\beqa
a_j(x)&=&{1\over 2\pi}
{\gamma\sin \gamma q_j\over \cosh\gamma x+\cos\gamma q_j},
~~\tilde{a}_j(\omega)={\sinh q_j\omega\over \sinh p_0\omega},\label{eq:ajx1}\\
q_j&\equiv&(-1)^i(p_i-(j-m_i)p_{i+1}),~~{\rm for}~~m_i\le j<m_{i+1},
\label{eq:ajx2}\enqa
where the $p_i,~~i=0,1,...,l$ are defined by
\beq
p_0=\pi/\gamma,~~ p_1=1,~~\nu_i=[p_{i-1}/p_i],~~p_i=p_{i-2}-p_{i-1}\nu_{i-1}.
\label{eq:ajx3}\enq
For the series $q_j$ we find the following relations
\beqa
q_j={1\over 2}[(1-2\delta_{m_i,j})q_{j-1}+q_{j+1}],~~{\rm for}~~
m_i\le j\le m_{i+1}-2,\nonum\\
q_j=(1-2\delta_{m_{i-1},j})q_{j-1}+q_{j+1},~~{\rm for}~~
 j= m_i-1,~~i<l,\nonum\\
q_0=p_0,~~n_{m_l}+n_{m_l-1}=0.
\enqa
The Fourier transform of $T_{j,k}(x)$ is given by
\beqa
&&\tilde{T}_{j,k}(\omega)=\tilde{T}_{k,j}(\omega)\nonum\\
&&=2\sign(q_j)\coth(p_{i+1}\omega){\sinh((p_0-|q_j|)\omega)
\sinh(q_k\omega)\over \sinh(p_0\omega)}\nonum\\
&&+\delta_{j,m_l-1}\delta_{k,m_l}-\delta_{j,k},
~~~{\rm for}~~j\le k,~~m_i<j\le m_{i+1}.~~~~~~~~\label{eq:tjk1}
\enqa
At $j=1$ we have
\beq
\tilde{T}_{1,k}(\omega)=\sign(q_1)
2\cosh \omega \tilde{a}_k(\omega)-\delta_{1,k}.
\label{eq:t1k}\enq
At $j=m_l$ we have
\beqa
&&\tilde{T}_{m_l,k}(\omega)=-\tilde{T}_{m_l-1,k}(\omega),\nonum\\
&&\tilde{T}_{m_l,m_l}(\omega)=-\tilde{T}_{m_l-1,m_l}(\omega)
={\sinh((p_0-2p_l)\omega)\over \sinh(p_0\omega)}.~~~~~~~~~~~~~~
\enqa
The energy and entropy per site are given by
\beqa
&&e=-({J\Delta\over 4}+h)+\sum^{m_l}_{j=1}\int^{\infty}_{-\infty}g_j(x)
\rho_j(x)dx,\nonum\\
&&g_j(x)\equiv  -{2\pi J \sin\gamma\over\gamma} a_j(x)+2n_jh,
\enqa
and
\beq
{\it s}=\int^\infty_{-\infty}
\rho_j\ln(1+{\rho_j^h\over \rho_j})+\rho_j^h
\ln(1+{\rho_j\over \rho_j^h})dx.
\enq
To minimize the free energy density $e-T{\it s}$ with respect to $\rho_j$, 
we have
$$\delta(e-T{\it s})=\sum_j\int^\infty_{-\infty} 
g_j(x)\delta\rho_j(x)-T\Bigl\{
\delta\rho_j\ln(1+{\rho_j^h\over \rho_j})+\delta\rho_j^h
\ln(1+{\rho_j\over \rho_j^h})\Bigr\}dx.$$
The variation of (\ref{eq:xylinear}) gives
$$\delta\rho_j^h=-\delta\rho_j-\sign(q_j)
\sum T_{jk}*\delta\rho_k.$$
Thus we obtain 
\beqa
&&\delta(e-T{\it s})=T\sum_j\int^\infty_{-\infty} dx\delta\rho_j(x)\nonum\\
&&\Bigl\{{g_j(x)\over T}-
\ln({\rho_j^h\over \rho_j})+\sum_k
\sign(q_k)T_{j,k}*\ln(1+{\rho_k\over \rho_k^h})
\Bigr\}.\nonum
\enqa
At the thermodynamic equilibrium one obtains  
the following non-linear equations determining  
$\eta_j(x)\equiv\rho_j^h(x)/\rho_j(x)$,
\beq
\ln\eta_j(x)=g_j(x)/T+\sum_{k=1}^{m_l}\sign(q_k)
T_{k,j}*\ln(1+\eta_k^{-1}(x)),
~~ j=1,...,m_l.
\label{eq:xynonl}\enq
The free energy is given as follows,
$$f=e-T{\it s}=-({J\Delta\over 4}+h)$$
$$+\sum^{m_l}_{j=1}\int^{\infty}_{-\infty}
\rho_j(x)[g_j(x)-T\ln\eta_j(x)]-T[\rho_j+\rho_j^h]\ln(1+\eta_j^{-1}).$$
If we substitute (\ref{eq:xynonl}) into the first bracket and 
(\ref{eq:xylinear}) into the second, the  
$T_{j,k}$ terms are cancelled and we get
\beq
f=-({J\Delta\over 4}+h)-T\sum_{j=1}^{m_l}\sign(q_j)\int^\infty_{-\infty}
a_j(x)\ln(1+\eta_j^{-1}(x))dx.
\label{eq:xyfree1}
\enq
If one uses the $j=1$ case of equation (\ref{eq:xynonl}) and (\ref{eq:t1k}), 
one obtains
\beqa
&&f=-{J\Delta\over 4}-\sign(q_1)
{2\pi J\sin\gamma\over \gamma}\int^\infty_{-\infty} a_1(x)s_1(x)dx\nonum\\
&&-T\int^\infty_{-\infty} s_1(x)\ln(1+\eta_1(x))dx.\label{eq:xyfree2}
\enqa
From equation (\ref{eq:ajx1}-\ref{eq:ajx3}) we get the following relations,
\beqa
&&a_j-s_i*((1-2\delta_{m_{i-1},j})a_{j-1}+a_{j+1})=0~~~\nonum\\
&&{\rm for}~~m_{i-1}\le j\le m_i-2,\nonum\\
&&a_{m_i-1}-(1-2\delta_{m_{i-1},m_i-1})s_i*a_{m_i-2}-d_i*a_{m_i-1}
-s_{i+1}*a_{m_i}=0  \nonum\\
&&{\rm for}~~i<l,\nonum\\
&&a_{m_l-1}(x)=-a_{m_l}(x)=s_l*a_{m_l-2},\label{eq:ajrel}
\enqa
where
\beqa
&&a_0(x)=\delta(x),\nonum\\
&&s_i(x)\equiv \int^\infty_{-\infty}{d\omega\over 4\pi}{e^{i\omega x}
\over \cosh(p_i\omega)}={1\over 4 p_i}{\rm sech}{\pi x\over 2 p_i},\nonum\\
&&d_i(x)\equiv \int^\infty_{-\infty}{d\omega\over 4\pi}{e^{i\omega x}
\cosh((p_i-p_{i+1})\omega)\over \cosh(p_i\omega)\cosh(p_{i+1}\omega)}.
\enqa
Using (\ref{eq:tjk1}) one can show the following relations,
\beqa
&&T_{j,k}-s_i*((1-2\delta_{m_{i-1},j})T_{j-1,k}+T_{j+1,k})\nonum\\
&&=(-1)^{i+1}
(\delta_{j-1,k}+\delta_{j+1,k})s_i,\nonum \\
&&~~{\rm for}~~
m_{i-1}\le j\le m_i-2,\nonum\\
&&T_{m_i-1,k}-(1-2\delta_{m_{i-1},m_i-1})s_i*T_{m_i-2,k}
-d_i*T_{m_i-1,k}\nonum\\
&&-s_{i+1}*
T_{m_i,k}=(-1)^{i+1}(\delta_{m_i-2,k}s_i+\delta_{m_i-1,k}d_i-\delta_{m_i,k}
s_{i+1}),\nonum\\
&&{\rm for}~~i=1,2,.,l-1,\nonum\\
&&T_{m_l-1,k}=-T_{m_l,k}=s_{l}*T_{m_l-2,k}+\sign(q_k)\delta_{m_l-2,k}s_l,
\label{eq:tjrel}\enqa
with $T_{0,k}=0$.
Using (\ref{eq:ajrel}) and (\ref{eq:tjrel}) 
one can rewrite (\ref{eq:xylinear}) as follows
\beqa
&&\rho_j+\rho_j^h=s_i*(\rho_{j-1}^h+\rho_{j+1}^h)~~{\rm for}~~
m_{i-1}\le j\le m_i-2,\nonumber\\
&&\rho_{m_i-1}+\rho_{m_i-1}^h=s_i*\rho_{m_i-2}^h+d_i*\rho_{m_i-1}^h
-s_{i+1}*\rho_{m_i}^h,\nonum\\
&&\rho_{m_l-1}+\rho_{m_l-1}^h=\rho_{m_l}+\rho_{m_l}^h=s_l*\rho_{m_l-1}^h,
\enqa
with $\rho_0^h=\delta(x)$. Equations (\ref{eq:xynonl}) are rewritten as 
\beqa
&&\ln(1+\eta_0)=-{2\pi J\sin\gamma\over \gamma T}\delta(x),\nonum \\
&&\ln \eta_j=(1-2\delta_{m_{i-1},j})s_i*\ln(1+\eta_{j-1})+s_i
*\ln(1+\eta_{j+1}),\nonum\\
~~&&{\rm for}~~m_{i-1}\le j\le m_i-2,j\ne m_l-2\nonum\\
&&\ln\eta_{m_i-1}=(1-2\delta_{m_{i-1},m_i-1})s_i*\ln(1+\eta_{m_i-2})\nonum\\
&&+d_i*\ln
(1+\eta_{m_i-1})+s_{i+1}*\ln(1+\eta_{m_i}),~~{\rm for}~~i<l\nonum\\
&&\ln\eta_{m_l-2}=(1-2\delta_{m_{l-1},m_l-2})s_l*\ln(1+\eta_{m_l-3})\nonum\\
&&+s_l*\ln((1+\eta_{m_l-1})(1+\eta^{-1}_{m_l})),\nonum\\
&&\ln\eta_{m_l-1}-y_lh/T=y_lh/T-\ln\eta_{m_l}\nonum\\
&&=s_l*\ln(1+\eta_{m_l-2}).
\label{eq:xynol2}
\enqa
Then if we write $\ln \kappa(x)=\ln\eta_{m_l-1}-y_lh/T$ we have integral 
equations with $m_l-1$ unknown functions
\beqa
&&\ln(1+\eta_0)=-{2\pi J\sin\gamma\over \gamma T}\delta(x),\nonum \\
&&\ln \eta_j=(1-2\delta_{m_{i-1},j})s_i*\ln(1+\eta_{j-1})+s_i
*\ln(1+\eta_{j+1}),\nonum\\
~~&&{\rm for}~~m_{i-1}\le j\le m_i-2,j\ne m_l-2\nonum\\
&&\ln\eta_{m_i-1}=(1-2\delta_{m_{i-1},m_i-1})s_i*\ln(1+\eta_{m_i-2})\nonum\\
&&+d_i*\ln
(1+\eta_{m_i-1})+s_{i+1}*\ln(1+\eta_{m_i}),~~{\rm for}~~i<l\nonum\\
&&\ln\eta_{m_l-2}=(1-2\delta_{m_{l-1},m_l-2})s_l*\ln(1+\eta_{m_l-3})\nonum\\
&&+s_l*\ln(1+2\cosh(y_lh/T)\kappa+\kappa^2),\nonum\\
&&\ln\kappa(x)=s_l*\ln(1+\eta_{m_l-2}).
\label{eq:xynol3}
\enqa
\index{Takahashi-Suzuki equation for XXZ model}

\section{Some special limits}
\subsection{$T\to \infty$ or $J\to {\it 0}$ limit}
In equations (\ref{eq:xynol3}), $\ln(1+\eta_0)$ becomes zero 
and $\eta_j(x)$ are all independent of $x$. This yields the following 
difference equation
\beqa
&&\eta_j^2=(1+\eta_{j-1})^{1-2\delta_{m_{i-1},j}}(1+\eta_{j+1})\nonum\\
&&~~{\rm for}~~m_{i-1}\le j\le m_i-2,j\ne m_l-2\nonum\\
&&\eta_{m_i-1}^2=(1+\eta_{m_i-2})^{1-2\delta_{m_{i-1},m_i-1}}
(1+\eta_{m_i-1})\ln(1+\eta_{m_i}),\nonum\\
&&~~{\rm for}~~i<l\nonum\\
&&\eta_{m_l-2}^2=(1+\eta_{m_l-3})^{1-2\delta_{m_{l-1},m_l-2}}
(1+2\cosh(y_lh/T)\kappa+\kappa^2),\nonum\\
&&\kappa^2=(1+\eta_{m_l-2}).
\enqa
\index{difference equation}
The solution of this set of equations is
\beqa
&&\eta_j=({\sinh(n_j+y_{i-1})h/T\over\sinh(y_{i-1}h/T)})^2-1\nonum\\
&&~~{\rm for}~~m_{i-1}< j\le m_i,~~j\le m_l-2,~~\nonum\\
&&\kappa={\sinh(n_{m_l-2}+y_l)h/T\over\sinh(y_lh/T)}.
\enqa
For $j=1$ we have $\eta_1=(2\cosh h/T)^2-1$. 
Substituting this into (\ref{eq:xyfree2}) we find the free energy 
\beq
f/T=-\ln(2\cosh h/T).
\enq
At $h=0$ this gives that entropy per site is $\ln2$, as it should be.

\subsection{$J>{\it 0},~~T\to {\it 0}$ limit}
We define $\epsilon_j(x)=T\ln\eta_j(x)$ and $\epsilon_j^+(x)=
T\ln(1+\eta_j(x))$. One can show that $\epsilon_j,~j\ge2$ is always positive. 
The equation (\ref{eq:xynonl}) gives
\beq
\epsilon_1(x)=-{2\pi J\sin\gamma\over \gamma}a(x,1)+2h
-\int^\infty_{-\infty} a(x-y,2)\epsilon^-_1(y)dy. 
\enq
If $\epsilon(x)<0$ at  $|x|<B$ and $\epsilon(x)>0$ at 
$|x|>B$, then one obtains a linear integral equation for $\rho_1(x)$,
\beq
\rho_1(x)+\int^B_{-B}a(x-y,2)\rho_1(y)dy=a(x,1),
\enq
where
\beq
a(x,n)\equiv {1\over 2\pi}{\gamma\sin n\gamma\over\cosh\gamma x-\cos n\gamma}.
\label{eq:xxzfre}
\enq


\chapter{Thermodynamics of the XYZ model}
\section{Bethe-ansatz equations for the XYZ model.}
Here we consider the symmetry of the following Hamiltonian
\beq
{\cal H}=-\sum_{l=1}^N J_xS^x_lS^x_{l+1}+J_yS^y_lS^y_{l+1}+J_xS^z_lS^z_{l+1}.
\label{eq:hamxyz2}\enq
We assume $N$ is even. By the transformation
$$U_2{\cal H}U_2^{-1}, ~~~U_2=\prod_{l=even}2S^z_l,$$ 
${\cal H}(J_x,J_y,J_z)\to {\cal H}(-J_x,-J_y,J_z)$.   In the same way
\par\noindent
${\cal H}(J_x,J_y,J_z)\to {\cal H}(J_x,-J_y,-J_z)$ and
${\cal H}(J_x,J_y,J_z)\to {\cal H}(J_x,-J_y,-J_z)$. Namely the energy 
spectrum of this Hamiltonian is unchanged for reversing signs of 
two $J_\alpha$'s. It is evident that the spectrum is unchanged for 
exchanging $J_\alpha$'s. Thus it is sufficient to treat only the case 
$1\ge J_y/J_z\ge |J_x|/J_z\ge 0$. Baxter solved the eight-vertex model 
and also the XYZ model (\cite{bax72a}, \cite{bax72b}). 
The Bethe ansatz equation for this model is
\beqa
&&\bigl({H_l(i\zeta(x_l+i))\over H_l(i\zeta(x_l-i))}\Bigr)^N=
-e^{-2\pi i\nu' /p_0}\prod_{j=1}^{N/2}
{H_l(i\zeta(x_l-x_j+2i))\over H_l(i\zeta(x_l-x_j-2i))},\nonum\\
&&\sum_l^{N/2}x_l=Q\nu'+ip_0\nu,~~Q=K(l')/\zeta,~~ p_0=K(l)/\zeta.\label{eq:xyzeqa3}
\enqa
Here the modulus $l$ and the parameter $\zeta$ are determined by  
$$l=\sqrt{J_z^2-J_y^2\over J_z^2-J_x^2},~~\cn (2\zeta,l)=-J_x/J_z.$$
There are $N/2$ rapidities. 
$H_l(x)$ is the Jacobian elliptic function defined by
$$H_l(x)\equiv 2\sum_{n=1}^\infty (-1)^{n+1}q^{n(n-1)+1/4}
\sin(2n-1){\pi x\over 2K},~~q\equiv\exp(-{\pi K(l')\over K(l)}).$$
This function has the following properties,
\beq
H_l(x)=-H_l(x+2K(l))=-qe^{i\pi x/K(l)}H_l(x+2iK(l')).
\enq
In the Bethe 
ansatz equation (\ref{eq:xxzarg3}), the function $\sinh{\gamma\over 2}x$ is 
merely replaced by the elliptic theta function. 
The energy is given by
\beq
E=-{NJ_z\over 4}[1-{\pi\sn2\zeta \over \zeta}({\bf a}(0,1)+{\bf a}(Q,1))]
-{J_z\pi\sn 2\zeta \over \zeta}
\sum_{l=1}^{N/2}{\bf a}(x_l,1),\label{eq:xyzeng5}
\enq
where
\beq
{\bf a}(x,l)\equiv{1\over 2\pi i}{d\over dx}\ln\Bigl({H_l(i\zeta(x+il))
\over H_l(i\zeta(x-il))}\Bigr).
\enq

\section{Strings and the thermodynamic Bethe ansatz equation for 
the XYZ model}
In the limit $l\to 0$ equation (\ref{eq:xyzeqa3}) becomes
\beq
\bigl({\sin(i\zeta(x_l+i))\over \sin(i\zeta(x_l-i))}\Bigr)^N=
-e^{-2\pi i\nu' /p_0}\prod_{j=1}^{N/2}
{\sin(i\zeta(x_l-x_j+2i))\over \sin(i\zeta(x_l-x_j-2i))}.
\enq
This equation is equivalent to (\ref{eq:xxzarg3}), if we assume 
that the $x_l$'s are finite and $\nu'=0$. $\zeta$ becomes $\gamma/2$ 
and $K_l$ becomes $\pi/2$. So it is natural to 
assume the same types of strings can be determined using $p_0=K_l/\zeta$,
\beqa
x_j&=&\alpha+(n+1-2j)i,\nonum\\
x_j&=&\alpha+(n+1-2j)i+p_0 i,~~Q\ge\alpha>-Q.\label{eq:string3}
\enqa
The solution becomes doubly periodic. 
So we consider the distribution of solutions at $-Q<\Re x\le Q$ and 
$-p_0<\Im x\le p_0$. It is expected that the same kind of strings 
appear in the case of XXZ model at $\pi/\gamma=p_0$. We can determine 
$n_j$'s and $q_j$'s via (\ref{eq:nj1}) and (\ref{eq:ajx2}),
\beqa
&&\{e_j(x_\alpha^j)\}^N=-\exp(-2\pi i \nu'/p_0)
\prod_{k=1}^{m_l}\prod_{\beta=1}^{M_k}
E_{jk}(x_\alpha^j-x_\beta^k),\nonum\\
&&\nu'={1\over Q}\sum_{j=1}^{m_l}\sum_{\alpha=1}^{M_j}n_jx_\alpha^j,
\label{eq:xyzsteq}
\enqa
where
\beqa
&&e_j(x)=g(x;n_j,v_j),\\
&&g(x;n,+)={H_l(i\zeta(x+in))\over H_l(i\zeta(x-in))}, \nonum\\
&&g(x;n,-)=-{H_l(K_l+i\zeta(x+in))\over H_l(K_l+i\zeta(x-in))},\\
&&E_{jk}(x)=\left\{\matrix{
g(x;2n_j,v_jv_k)\prod_{l=1}^{n_j-1}g^2(x;2l,v_jv_k)~~{\rm for}~~n_j=n_k,\cr
g(x;(n_j+n_k),v_jv_k)g(x;|n_j-n_k|,v_jv_k)\cr
\times \prod_{l=1}^{Min(n_j,n_k)-1}g^2(x;|n_j-n_k|+2l,v_jv_k)\cr
~~{\rm for}~~n_j\ne n_k,}\right.
\enqa 
Taking the logarithm of (\ref{eq:xyzsteq}) we have
\beq
N\theta_j(x_\alpha^j)=2\pi I_\alpha^j-2\pi \nu'/p_0 
+\sum_{k=1}^{m_l}\sum_{\beta=1}^{M_k}
\Theta_{jk}(x_\alpha^j-x_\beta^k),~~\alpha=1,2,...,M_j.
\enq
Here 
\beqa
&&\theta_j(x)={\bf f}(x;n_j,v_j),
~~\Theta_{jk}(x)={\bf f}(x;|n_j-n_k|,v_jv_k)+\nonum\\
&&{\bf f}(x;n_j+n_k,v_jv_k)
+2\sum_{i=1}^{Min(n_j,n_k)-1}
{\bf f}(x;|n_j-n_k|+2i,v_jv_k),\label{eq:xyzeng6}
\enqa
and ${\bf f}(x,n,v)$ is defined by
\[{\bf f}(x,n,v)=f(x,n,v)+\sum_{l=1}^\infty f(x-2lQ,n,v)+f(x+2lQ,n,v).\]
$f(x,n,v)$ was defined in (\ref{eq:funcf}). 
An eigenstate should be identified by the set of quantum numbers 
$I_j^\alpha$. 
From (\ref{eq:xyzeng5})
the energy must be
\beqa
&&E=-NJ_zR
-{J_z\pi\sn 2\zeta \over \zeta}
\sum_{l=1}^{N/2}{\bf a}(x_l,1)
=-NJ_zR
-{J_z\pi\sn 2\zeta \over \zeta}\sum_{j=1}^{m_l}
\sum_{\alpha=1}^{M_j}{\bf a}_j(x_\alpha),\nonum\\
&&R\equiv{1\over 4}[1-{\pi\sn2\zeta \over \zeta}({\bf a}(0,1)+{\bf a}(Q,1))],
\nonum\\
&&{\bf a}_j(x)\equiv {1\over 2Q}\Bigl[{q_j\over p_0}+2\sum_{l=1}^\infty
{\sinh (q_j\pi l/Q)\over \sinh (p_0\pi l/Q)}\cos(\pi jx/Q)\Bigr],
\label{eq:xyzen1}
\enqa
where $q_j$ was defined in (\ref{eq:ajx2})
The number of zeros must be $N/2$, so
\beq
N/2=\sum_{j=1}^{m_l}n_jM_j.\label{eq:xyzmag}
\enq 
Thus the energy per site is given by
\beq
e=-J_zR
-{J_z\pi\sn 2\zeta \over \zeta}\sum_{j=1}^{m_l}\int_{-Q}^{Q}
{\bf a}_j(x)\rho_j(x)dx.\label{eq:xyzen2}
\enq
The entropy per site is
\beq
{\it s}=\sum_{j=1}^{m_l}\int^Q_{-Q}
\rho_j\ln(1+{\rho_j^h\over \rho_j})+\rho_j^h
\ln(1+{\rho_j\over \rho_j^h})dx.
\enq 
From (\ref{eq:xyzeng6}) we have the relation between 
$\rho_j(x)$ and $\rho_j^h(x)$,
\beq
{\bf a}_j(x)=\sign(q_j)(\rho_j(x)+\rho_j^h(x))
+\sum_{k=1}^{m_l}{\bf T}_{j,k}*\rho_k(x).
\label{eq:xyzlinear}
\enq
Moreover, from (\ref{eq:xyzmag}),
\beq
{1\over 2}=m\equiv\sum_{j=1}^{m_l}n_j\int^Q_{-Q}\rho_j(x)dx.\label{eq:xyzmag2}
\enq
Next we need a Lagrange multiplier to guarantee the 
condition (\ref{eq:xyzmag2}). \index{Lagrange multiplier}
One should minimize $e-T{\it s}+2h m$ under conditions (\ref{eq:xyzlinear}), 
and after that the multiplier $h$ should be chosen so that 
(\ref{eq:xyzmag2}) is satisfied. Just in the same way as before 
we get the integral equations for 
$\eta_j(x)={\rho^h_j(x)/\rho_j(x)}$, 
\beq
\ln\eta_j(x)={\bf g}_j(x)/T+\sum_{k=1}^{m_l}\sign(q_k)
{\bf T}_{k,j}*\ln(1+\eta_k^{-1}(x)),
~~ j=1,...,m_l.
\label{eq:xyznonl}\enq
Here $*$, ${\bf g}_j(x)$ and ${\bf T}_{j,k}(x)$ are
$$f*g(x)=\int^Q_{-Q} f(x-y)g(y)dy,$$
$${\bf g}_j(x)\equiv -{J_z\pi \sn 2\zeta\over \zeta}{\bf a}_j(x)+2n_jh,$$
$${\bf T}_{j,k}(x)={1\over 2Q}\sum_{l=-\infty}^\infty e^{i\pi nx/Q}\tilde{T}
_{j,k}({\pi l\over Q})=\sum_{l=-\infty}^{\infty} T_{j,k}(x-2lQ).$$
The quantity $g\equiv e-T{\it s}+2hm$ is given as follows
\index{block tri-diagonal equation}
\beqa
&&g(J_z,T,h)=-J_zR\nonum\\
&&+\sum^{m_l}_{j=1}\int^{Q}_{-Q}
\rho_j(x)\Bigl[{\bf g}_j(x)-T\ln\eta_j(x)\Bigr]
-T\Bigl[\rho_j+\rho_j^h\Bigr]\ln(1+\eta_j^{-1})dx
\nonum\\
&&=-J_zR-T\sum_{j=1}^{m_l}\sign(q_j)\int^Q_{-Q}
{\bf a}_j(x)\ln(1+\eta_j^{-1}(x))dx.
\label{eq:xyzfree1}
\enqa
Corresponding to (\ref{eq:xyfree2}) this is 
\beqa
&&g=-{J_z R}+h-\sign(q_1)
{\pi J_z\sn 2\zeta\over \zeta}\int^Q_{-Q} {\bf a}_1(x){\bf s}_1(x)dx\nonum\\
&&-T\int^Q_{-Q} {\bf s}_1(x)\ln(1+\eta_1(x))dx.\nonum\\
\label{eq:xyzfree2x}
\enqa
Then $m$ should be determinded by
\beq
m={1\over 2}{\partial g\over \partial h}={1\over 2}
-{1\over 2}\int^Q_{-Q}{\bf s}_1(x)
(1+\eta_1(x))^{-1}{\partial \eta_1(x)\over \partial h}dx.\label{eq:mdet}
\enq
The equation (\ref{eq:xyznonl}) is also equivalent to the following 
block tridiagonal equations,
\beqa
&&\ln(1+\eta_0)=-{\pi J_z\sn 2\zeta\over \zeta T}\delta(x),\nonum \\
&&\ln \eta_j=(1-2\delta_{m_{i-1},j}){\bf s}_i*\ln(1+\eta_{j-1})+{\bf s}_i
*\ln(1+\eta_{j+1})\nonum\\
~~&&{\rm for}~~m_{i-1}\le j\le m_i-2,j\ne m_l-2,\nonum\\
&&\ln\eta_{m_i-1}=(1-2\delta_{m_{i-1},m_i-1}){\bf s}_i*\ln(1+\eta_{m_i-2})\nonum\\
&&+{\bf d}_i*\ln
(1+\eta_{m_i-1})+{\bf s}_{i+1}*\ln(1+\eta_{m_i})~~{\rm for}~~i<l,\nonum\\
&&\ln\eta_{m_l-2}=(1-2\delta_{m_{l-1},m_l-2}){\bf s}_l*\ln(1+\eta_{m_l-3})\nonum\\
&&+{\bf s}_l*\ln(1+2\cosh (y_lh/T) \kappa+\kappa^2),\nonum\\
&&\ln\kappa={\bf s}_l*\ln(1+\eta_{m_l-2}).
\label{eq:xyznon2}
\enqa
In this equation parameter $h$ appears only in $\cosh (y_lh/T)$ term. 
So $\eta_1(x,h)$ is even function of $h$ and 
${\partial \eta_1(x)\over \partial h}|_{h=0}=0$.  Using (\ref{eq:mdet}) 
we find $m=1/2$ and 
the condition (\ref{eq:xyzmag2}) is satisfied at $h=0$. 
Thus equation (\ref{eq:xyznon2}) becomes
\beqa
&&\ln(1+\eta_0)=-{\pi J_z\sn 2\zeta\over \zeta T}\delta(x),\nonum \\
&&\ln \eta_j=(1-2\delta_{m_{i-1},j}){\bf s}_i*\ln(1+\eta_{j-1})+{\bf s}_i
*\ln(1+\eta_{j+1})\nonum\\
~~&&{\rm for}~~m_{i-1}\le j\le m_i-2,j\ne m_l-2,\nonum\\
&&\ln\eta_{m_i-1}=(1-2\delta_{m_{i-1},m_i-1}){\bf s}_i*\ln(1+\eta_{m_i-2})\nonum\\
&&+{\bf d}_i*\ln
(1+\eta_{m_i-1})+{\bf s}_{i+1}*\ln(1+\eta_{m_i})~~{\rm for}~~i<l,\nonum\\
&&\ln\eta_{m_l-2}=(1-2\delta_{m_{l-1},m_l-2}){\bf s}_l*\ln(1+\eta_{m_l-3})\nonum\\
&&+2{\bf s}_l*\ln(1+\kappa),\nonum\\
&&\ln\kappa={\bf s}_l*\ln(1+\eta_{m_l-2}).
\label{eq:xyznon3}
\enqa
\index{Takahashi-Suzuki equation for XYZ model}
Corresponding to (\ref{eq:xyfree2}) the free energy is 
\beqa
&&f=-{J_z R}-\sign(q_1)
{\pi J_z\sn 2\zeta\over \zeta}\int^Q_{-Q} {\bf a}_1(x){\bf s}_1(x)dx\nonum\\
&&-T\int^Q_{-Q} {\bf s}_1(x)\ln(1+\eta_1(x))dx.\nonum\\
\label{eq:xyzfree2}
\enqa
We can calculate free energy of XYZ model in zero external field. 

\chapter{Numerical calculation and recent developments}
In this lecture we restricted ourselves to the unnested Bethe ansatz. 
Fermions with $\delta$-function interactions and the Hubbard model 
belong to the nested Bethe ansatz which has several kinds of rapidities. 
For details see (\cite{tak71b}, \cite{lai71}, \cite{tak72}).
But the method of derivation is essentially the same as the unnested cases.

The equations introduced in this lecture note have been solved numerically. 
One can calculate the specific heat,  magnetic susceptibility, 
magnetization curve for the XXZ model, and the specific heat for the XYZ model 
at $p_0=2,3$ (\cite{tak74a}).  
If we increase $p_0$ the problem approaches to the $\Delta=-1$ case. 
For the analysis of thermodynamic quantities not at low temperature, 
we can use the high-temperature expansion method or 
exact diagonalization. 
For the investigation of low-temperature thermodynamics of solvable 
models the Bethe ansatz method is the only way. For the spin 1/2 
ferromagnetic XXX 
chain, the susceptibility diverges as $T^{-\gamma}$ and the 
specific heat behaves as $T^{-\alpha}$. The estimations of exponent 
$\gamma$ had been done 
by many authors. Baker et. al. estimated $\gamma=1.66$ using high 
temperature series expansions (\cite{baker64}). Lyklema obtained $\gamma=1.75$
using quantum Monte Carlo calculations (\cite{lyk83}). 
By the numerical calculation of thermodynamic Bethe 
ansatz equations it was established that $\gamma=2$ and $\alpha=1/2$
(\cite{tak85}, \cite{yam86}, \cite{schlot85}).  \index{low-temperature thermodynamics} This investigation continued to the spin-wave theory 
for low-dimensional magnets (\cite{tak86}, \cite{tak87}, \cite{tak89}). 

In many cases of one-dimensional quantum systems, one can define the quantum 
transfer matrix. The largest eigenvalue of this matrix gives the free energy.  
The ratio of the largest and the second largest eigenvalues gives the 
correlation length. This method was developed mainly by Japanese theorists within the last ten years. 
The XYZ model and the Hubbard model were investigated 
by this method, for other solvable models this method is expected to be 
applicable. Logarithmic anomalies in the low-temperature susceptibility 
of the XXX antiferromagnet were found by the numerical calculation of 
Bethe ansatz equations for the largest eigenvalue of the quantum transfer 
matrix (\cite{egg}).

%

\printindex
\end{document}